\documentclass[preprint,letterpaper]{revtex4}
\usepackage{comment}
\usepackage{multirow}
\usepackage{array}
\usepackage{bm}
\usepackage{amsmath}
\usepackage{longtable}
\usepackage{color}
\usepackage{ textcomp }
\usepackage{diagbox}
\usepackage[utf8]{inputenc}
\usepackage{bm}
\usepackage{graphicx}
\usepackage{tikz}
\usetikzlibrary{quantikz}
\usepackage{xcolor} 

\begin{document}

\title{Simulating Energy Transfer in Molecular Systems with Digital Quantum Computers}

\author{Chee-Kong Lee}
\email{cheekonglee@tencent.com}
\affiliation{Tencent America, Palo Alto, CA 94306, United States}
\author{Jonathan Wei Zhong Lau}
\email{e0032323@u.nus.edu}
\affiliation{Centre for Quantum Technologies, National University of Singapore, 117543, Singapore}
\author{Liang Shi}
\email{lshi4@ucmerced.edu}
\affiliation{Chemistry and Chemical Biology, University of California, Merced, California 95343, United States}
\author{Leong Chuan Kwek}
\email{cqtklc@nus.edu.sg}
\affiliation{Centre for Quantum Technologies, National University of Singapore, 117543, Singapore}
\affiliation{National Institute of Education, Nanyang Technological University, 1 Nanyang Walk, Singapore 637616}
\affiliation{MajuLab, CNRS-UNS-NUS-NTU International Joint Research Unit, UMI 3654, Singapore}

\begin{abstract}
Quantum computers have the potential to simulate chemical systems beyond the capability of classical computers. Recent developments in hybrid quantum-classical approaches enable the determinations of the ground or low energy states of molecular systems. Here, we extend near-term quantum simulations of chemistry to time-dependent processes by simulating energy transfer in organic semiconducting molecules. 
We developed a multi-scale modeling workflow that combines conventional molecular dynamics and quantum chemistry simulations with hybrid variational quantum algorithm to compute the exciton dynamics in both the single excitation subspace (i.e. Frenkel Hamiltonian) and the full-Hilbert space (i.e. multi-exciton) regimes.
Our numerical examples demonstrate the feasibility of our approach, and simulations on IBM Q devices capture the qualitative behaviors of exciton dynamics, but with considerable errors. We present an error mitigation technique that combines experimental results from the variational and Trotter algorithms, and obtain significantly improved quantum dynamics. Our approach opens up new opportunities for modeling quantum dynamics in chemical, biological and material systems with quantum computers.
\end{abstract}

\clearpage

\maketitle

\section{Introduction}
\begin{figure}
  \includegraphics[width=.95\textwidth]{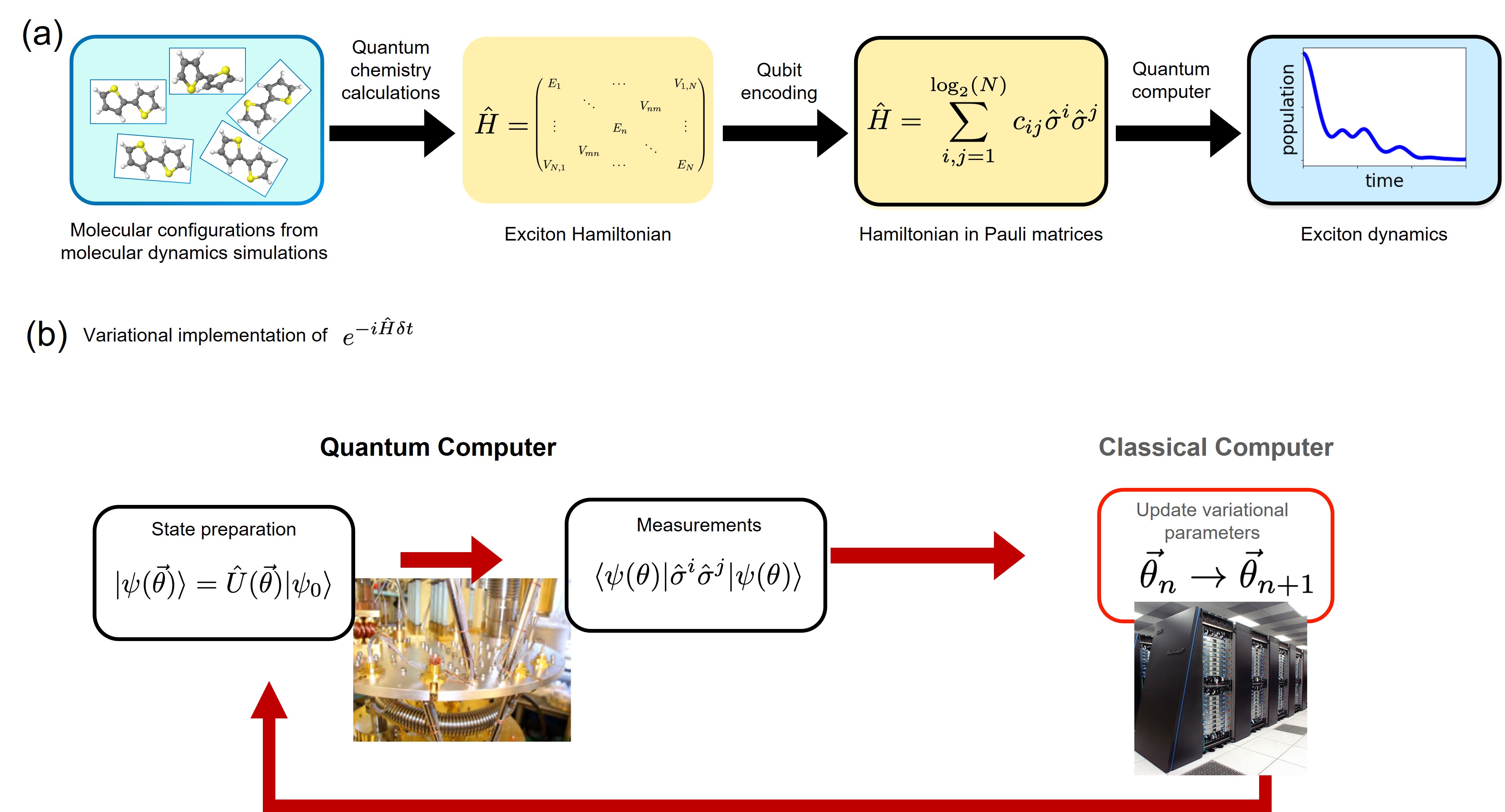}
  \caption{(a) A schematic of the setup in this work. The molecular configurations are obtained from molecular dynamics simulations. From these molecular configurations, quantum chemistry calculations are utilized to obtain the exciton Hamiltonian. An $N$-molecule exciton Hamiltonian is either  encoded in $\log_2(N)$ qubits within the single exciton approximation or $N$ qubits within the full-Hilbert space (i.e. multi-exciton) treatment. Finally we simulate the quantum dynamics of the exciton system with a digital quantum computer. 
  (b) A variational quantum algorithm is used to simulate the time evolution of excitonic system. The wavefunction at time $t$ is represented by a parametrized circuit ansatz, $\ket{\Psi(t)} \approx  \ket{\psi(\vec{\theta}(t))}$. The variational parameters are updated iteratively with a classical computer based on the measurement outputs from the quantum computer. The updated parameters are then sent to quantum computer for preparation of a new quantum state.}
  \label{fig:schematic}
\end{figure}

Quantum simulation of chemical systems is one of the most promising and anticipated applications for quantum computers\cite{Lanyon2010, Aspuru-Guzik2005, Kassal2011, McArdle2020, Cao2019, Bauer2020, MacDonell2021}.
Classical simulation of quantum systems becomes exponentially 
complex with an increasing number of degrees of freedom. Even for a modest-sized molecule comprising of tens of atoms, accurate  simulation on a classical computer remains challenging.
A quantum computer reputedly allows us to tackle some of these classically intractable computational problems more efficiently. Such computations enable us to design better compounds for chemistry, materials science and biology.
In recent years, there are numerous experimental and theoretical works that focus on the determination of the ground or low-energy eigenstates of molecular systems with near-term digital quantum computers~\cite{Cerezo2020, Peruzzo2014, Kandala2017, Kandala2019, OMalley2016, Motta2020, AIQuantum2020, hempel2018quantum, Parrish2019b, Takeshita2020, OBrien2019, Grimsley2019, Nam2020, McCaskey2019, Hsieh2021}. 
However, many phenomena in chemistry are dynamical and require a time-dependent solution, e.g. energy transfer, chemical reactions, and spectroscopy~\cite{mukamel95, nitzan2013chemical, Gatti2017}.
Understanding these processes would therefore require the simulation of the time evolution of the relevant quantum systems. 
This task is typically more difficult than the computation of static ground state properties since it often involves the participation of all quantum states.
Particularly, the no-fast-forwarding theorem states that simulating the dynamics of a quantum system for time $t$ typically requires $O(t)$ gates; in other words, a generic Hamiltonian evolution cannot be achieved in sublinear time~\cite{Berry2007}. 
Despite its relevance and importance, research on simulating time-dependent processes in chemical systems with noisy intermediate-scale quantum (NISQ) devices remains limited~\cite{Heya2019, Zhang2020, Ollitrault2020, Lee2021}.

While there are a number of quantum algorithms to simulate the dynamics of quantum many-body systems, such as those based on Trotterization~\cite{Lloyd1996}, linear combination of unitaries~\cite{Childs2012}, quantum signal processing~\cite{Low2017} and qubitization~\cite{Low2019}, these algorithms require long circuit depth and are therefore not suitable for NISQ devices due to the accumulation of hardware errors.
Hence we adopt a variational quantum algorithm (VQA) based on the McLachlan's principle to simulate these time-dependent processes in chemical systems~\cite{Li2017, Endo2020, Chen2020}.
Within the VQA, the time-dependent wavefunction is represented by a parametrized quantum state $\ket{\psi(\vec{\theta})}$ that can be prepared efficiently in a quantum computer. The variational parameters are updated iteratively by the classical computer via the optimization of the time-dependent Schrodinger equation.
As a result of the integration with classical computer, the VQA algorithm can be implemented with quantum circuits of much shallower depth compared with other conventional algorithms for Hamiltonian simulations.

 
For noisy quantum systems, fault-tolerant quantum computing using quantum error correction codes has been proposed as a means to assuage the effects of errors. Quantum information is encoded in logical qubits each comprising of many physical qubits, and the computation error can be reduced to arbitrarily small value provided that the error of each physical qubit is smaller than a certain threshold value. For NISQ devices, the number of physical qubits is likely to be restricted.
Thus, before the realization of quantum error correction, error mitigation in NISQ devices has been proposed as an alternative to fault-tolerant quantum computing.
Several error mitigation techniques have since been developed, e.g. error extrapolation~\cite{Li2017}, quasi-probability method~\cite{Temme2017}, quantum subspace expansion~\cite{McClean2017} and machine learning~\cite{Kim2020}. 
In this work, we demonstrate a new error mitigation method for quantum dynamics by combining the experimental results from the VQA and a Trotterization scheme.
This error mitigation method is motivated by the intuition that Trotterized quantum simulation is accurate in short time even in noisy quantum circuits, therefore useful information can be extracted from the short-time Trotter dynamics to mitigate errors in the VQA. 
We show that such error mitigation method significantly improves the accuracy of quantum dynamics simulation by performing experiments on IBM quantum computers. 
To the best of our knowledge, it is the first demonstration of performing error mitigation for quantum dynamics simulation that combines the results of a hybrid variational algorithm and a fault tolerant algorithm.

 We demonstrate the feasibility of using a digital quantum computers to simulate quantum dynamics in chemical systems by studying the energy transfer process. To be exact, we consider excitonic energy transfer in organic semiconducting molecules. An exciton is a quasi-particle excitation consisting of a bound pair of electron and hole
that mediates the energy transport in a wide range of systems ranging from photosynthetic light harvesting systems, organic light-emitting diodes (OLEDs) to organic photovoltaics (OPVs)~\cite{Ostroverkhova2016, Mikhnenko2015, Deotare2015}.
Understanding exciton transfer in experimentally relevant systems is thus vital for various technological applications, e.g. designing more efficient and robust OPVs and OLEDs.
However, quantum mechanical simulation of the exciton dynamics in molecular systems can be challenging due to the presence of disorder (both static and dynamical) and the large number of excitonic sub-units involved. Quantum computers could potentially overcome these computational challenges. 

We aim to study chemical systems with direct relevance in chemistry and practical applications. Thus we choose bi-thiophene (T2) as our test system. Bi-thiophene is  electro-deposited on indium tin oxide (ITO) substrates for the fabrication of electrochromic devices. It can also be used in the formation of electrode material for the development of supercapacitors.  T2 molecule represents the minimum model within the family of thiophene-based polymers, one of the most studied semiconducting organic materials, and it has important applications in thin film technology, such as field-effect transistors~\cite{Perepichka2009}.

To include the effects of complex environment on quantum dynamics, we employ multi-scale modelling by performing molecular dynamics (MD) and quantum chemistry calculations to extract the input Hamiltonian for our quantum computer simulations.
The setup of this work is shown in Fig.~\ref{fig:schematic}a. We first perform classical MD simulation to obtain the molecular configurations of T2 molecules. For each MD snapshot, quantum chemistry calculations are used to obtain the instantaneous single-molecule excited-state energies and inter-molecular couplings. Once all the energies and couplings along the MD trajectory are computed, the time-dependent exciton Hamiltonian can then be constructed.  

We first review the time-dependent VQA in Sec.~\ref{sec:vqa}. 
We then present the details of the exciton model with the single excitation approximation (i.e. Frenkel exciton model) in Sec.~\ref{sec:exciton} and with the full-Hilbert space treatment (i.e. multi-exciton model) in Sec.~\ref{sec:full_hilbert}.
In both regimes, the excitonic wavefunction is evolved according to the encoded Hamiltonian using the VQA algorithm, as illustrated in Fig.~\ref{fig:schematic}b.
Specifically, we numerically simulate the exciton dynamics in a molecular crystal of T2 molecules, and observe good agreement between the VQA results and exact calculations.
We then assess the capability of current quantum devices in simulating exciton dynamics in Section \ref{sec:experiment} by implementing the VQA on IBM Q quantum devices. Despite that only short circuit depth is needed, the VQA only qualitatively captures the behaviors of the true exciton dynamics, with some considerable quantitative discrepancy.
Finally in Sec.~\ref{sec:mitigation}, we discuss the source of errors and propose a new error mitigation technique by combining the VQA result with that from a standard Trotter scheme. Motivated by the insight that the short-time dynamics from the Trotter method is accurate, we advocate correcting the VQA simulation results with the information extracted from the short-time Trotter dynamics. We demonstrate that this error-mitigation technique significantly improves the accuracy of the VQA simulation. 

\section{Variational Quantum Algorithm (VQA) for Quantum Dynamics}\label{sec:vqa}
To simulate the exciton dynamics, we adopt the time-dependent VQA introduced in Ref. \cite{Li2017, Endo2020, Chen2020}.
In this VQA, the time-dependent quantum state, $\ket{\Psi(t)}$, is approximated by a parametrized quantum state, $\ket{\psi(\theta(t))}$, i.e.
$\ket{\Psi(t)} = \hat T \mbox{e}^{-i \int^t_0 \hat{H}(t') dt' } \ket{\Psi_0} \approx \ket{\psi(\vec \theta(t))}$ where 
$\vec \theta(t) = [\theta_1(t), \theta_2(t), \theta_3(t) ...] $ denotes the variational parameters at time $t$ and $\hat{T}$ is the time-ordering operator. According to McLachlan’s principle, the equation of motion for the variational parameters is obtained by minimizing the quantity $\|\Big (i\frac{\partial}{\partial t} - \hat H(t) \Big) \ket{\psi(\theta}\|$ which results in
\begin{eqnarray} \label{eq:EOM_theta}
\vec \theta(t + \delta t) = \vec \theta(t) +  \dot{\vec \theta} (t) \delta t; \,\,\, \dot{\vec \theta} (t) = \hat M^{-1} \vec{V}, 
\end{eqnarray}
where the matrix elements of $\hat M$ and $\vec V$ are
\begin{eqnarray} \label{eq:EOM}
\hat M_{kl} =  \mbox{Re} \left\langle \frac{\partial \psi(\vec{\theta})}{\partial \theta_k} \right\vert \left. \frac{\partial \psi(\vec{\theta})}{\partial \theta_l}
\right\rangle; 
\vec V_{k} =  \mbox{Im} \left\langle  \psi(\vec{\theta}) \right\vert \hat H \left\vert \frac{\partial \psi(\vec{\theta})} {\partial \theta_k }
\right\rangle.  
\end{eqnarray}

The accuracy of VQA depends crucially on the choice of the wavefunction ansatz. Given the complexity of the encoded Hamiltonian, we need a powerful ansatz capable of accurately capturing the exciton dynamics, yet it can be efficiently implemented in a quantum computer. Here we consider an ansatz of the form
\begin{eqnarray} \label{eqn:wfn_ansatz}
\ket{\psi(\theta)} = \hat{U}(\vec \theta)\ket{\psi_0}  = \prod_{k} \hat{U}_k(\theta_k) \ket{\psi_0}  =  \prod_{k} \mbox{e}^{i \theta_k \hat{R}_k} \ket{\psi_0}. 
\end{eqnarray}
where $\ket{\psi_0}$ is the initial state of the wavefunction and $\hat{R}_k$ is some Pauli strings. In the numerical calculations in the following sections, we consider all combinations of single and 2-qubit rotations, i.e. $\hat{R}_k \in \{\hat \sigma^m_a, \hat \sigma^m_a \hat \sigma^n_{b}\}$ for $a,b = x,y,z$ and all qubit combinations. 
Thus the circuit depth for state preparation is in the order of $O(L^2)$ where $L$ is the number of qubits.
The above ansatz is similar to the popular Hamiltonian ansatz~\cite{Wecker2015} in which $\ket{\psi(\theta)} = \prod_{l } [\prod_{j} \mbox{e}^{i \theta_{j,l} h_j}] \ket{\psi_0}$ for a given Hamiltonian $\hat H = \sum_j c_j \hat h_j$ and $l$ represents the number of layers.
Similarly, he expressive power of the ansatz in Eq.~(\ref{eqn:wfn_ansatz}) can be systematically improved by adding more layers. 
It is worth noting that the multi-scale framework developed in this work is not limited to this particular wavefunction ansatz, other forms of ansatz for quantum dynamics can be adopted in a straightforward manner.~\cite{Yao2021, Lee2020a} 

\begin{figure}
    \centering
    (a)
    \begin{tikzcd}
    \lstick{$\frac{\ket{0}+e^{i\phi}\ket{1}}{\sqrt{2}}$}&\qw&\gate{X}&\qw&\ctrl{1}&\qw&\gate{X}&\qw&\ctrl{1}&\gate{H}&\meter{}\\
    \lstick{$\ket{\psi_0}$}&\gate{U_1}&\gate{\dots}&\gate{U_k}&\gate{R_k}&\gate{U_{k+1}}&\gate{\dots}&\gate{U_l}&\gate{R_l}&\qw&\qw
    \end{tikzcd}\\
    (b)
    \begin{tikzcd}
    \lstick{$\frac{\ket{0}+e^{i\phi}\ket{1}}{\sqrt{2}}$}&\qw&\gate{X}&\qw&\ctrl{1}&\qw&\gate{X}&\qw&\ctrl{1}&\gate{H}&\meter{}\\
    \lstick{$\ket{\psi_0}$}&\gate{U_1}&\gate{\dots}&\gate{U_k}&\gate{R_k}&\gate{U_{k+1}}&\gate{\dots}&\gate{U_L}&\gate{h_j}&\qw&\qw
    \end{tikzcd}
    \caption{Quantum circuits to compute the matrix elements of (a) $\hat M$ and (b) $\vec V$ in Eq.~(\ref{eqn:circuit}). The ancillary qubit is initialized in state $\frac{\ket{0} + e^{i\phi} \ket{1} }{\sqrt{2}}$. The phase factor $\phi$ is set to be $0$ or $\pi/2$ in order to measure the real and imaginary components of the expectation values, respectively.}
    \label{fig:circuit}
\end{figure}
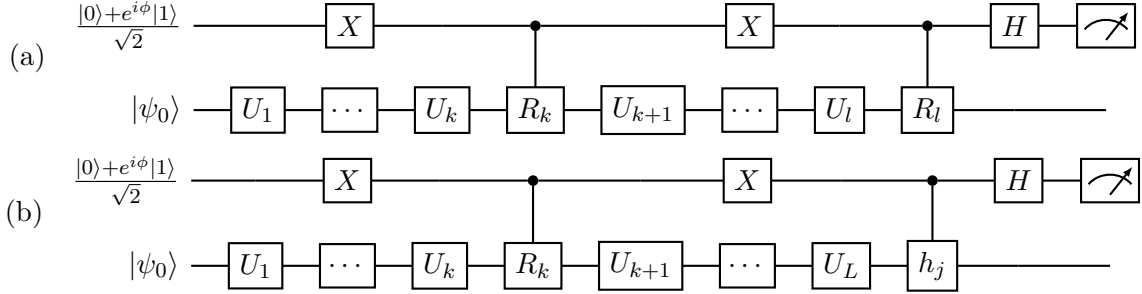

We next show that the matrix elements of $\hat M$ and $\vec V$ can be efficiently measured in a quantum computer.
We note that the derivative of each term in Eq.~(\ref{eqn:wfn_ansatz}) is written as $\frac{\partial \hat{U}_k(\theta_k)}{\partial \theta_k} = i \hat{R}_k \hat{U}_k(\theta_k)$, therefore the matrix elements of $\hat M$ and $\vec V$ are (assuming $k<l$)
\begin{eqnarray} \label{eqn:circuit}
\hat M_{kl} &=&   \mbox{Re} \big(  \bra{\psi_0} \hat U_1^\dagger  ...\hat U_k^\dagger \hat R_k^\dagger ... \hat U_L^\dagger
                        \hat U_L^ ... \hat R_l \hat U_l ... \hat U_1  \ket{\psi_0} \big) \\
       &=& \mbox{Re} \big( \bra{\psi_0} \hat U_1^\dagger  ...\hat U_k^\dagger \hat R_k^\dagger \hat U_{k+1}^\dagger  ...\hat U_l^\dagger \hat R_l \hat U_l ... \hat U_1  \ket{\psi_0} \big) ,  \nonumber \\
\vec V_{k}   &=&\mbox{Im} \big(  i \sum_j c_j \bra{\psi_0} \hat U_1^\dagger  ...\hat U_L^\dagger \hat h_j
                        \hat U_L^ ... \hat R_k \hat U_k ... \hat U_1  \ket{\psi_0} \big),   \nonumber
\end{eqnarray}
where we have expressed the Hamiltonian as $\hat H = \sum_j c_j \hat{h}_j$. The quantities in Eq.~(\ref{eqn:circuit}) are obtained in a quantum circuit via the Hadamard test, and the structures of the circuits are shown in Fig.~\ref{fig:circuit}.

It is worth nothing that the no-fast-forwarding theorem states that simulating the dynamics of a quantum system for time $t$ requires $O(t)$ quantum resources from the complexity point of view~\cite{Berry2007}. In our case, the time-dependent VQA breaks down the simulation into many small quantum calculations of fixed depth. It is the total number of these calculations that scales with $O(t)$ instead of the circuit depth.


\section{Frenkel Exciton Hamiltonian} \label{sec:exciton}
We adopt an {\it ab-initio} exciton model where the electronic excitation in a multi-chromophoric system is expressed in a basis of localized Frenkel excitations with input from MD simulation and quantum chemical calculations.~\cite{Shi2018, Sisto2014}
The resulting (Frenkel) exciton Hamiltonian describes a system of $N$ excitable subunits (also called sites or chromophores) that can each host a localized exciton. The ground electronic state of the system is denoted as $\ket{0}$, the wavefunction for the localized exciton is $\ket{m} = \hat{e}^\dagger_m \ket{0}$, where the creation operator $\hat{e}^\dagger_m$ generates an exciton localized on the $m$-th site. The form of the creation operator depends on the selected excited-state electronic-structure method. In the basis of these wavefunctions, the exciton Hamiltonian is written as
\begin{equation}
    \label{eqn:frenkel}
    \hat{H}_{\text{exciton}} = \sum_{m} E_m(t) \ket{m}\bra{m} + \sum_{m\neq n} V_{mn}(t) \ket{m}\bra{n},
\end{equation}
where $E_{m}$ is the excitation energy of molecule $m$ and $V_{mn}$ is the coupling between local excitations on molecules $m$ and $n$. 

The time dependence of the excitation energies and excitonic couplings in the exciton Hamiltonian originates from the modulation of the electronic excitation by nuclear motion, sometimes termed exciton-phonon couplings, which heavily influences exciton transport~\cite{Mohseni2008, Rebentrost2009, Chin2013}. There are multiple approaches to model the exciton-phonon couplings, e.g. phenomenologically through Gaussian white noise or a bath of quantum harmonic oscillators. 
Here, we adopt an all-atom approach by combining classical MD simulations and time-dependent density functional theory (TDDFT) quantum chemistry calculations.
In this approach, the time-dependent fluctuation of the exciton Hamiltonian is provided by fully atomistic simulations without resort to phenomenological models. 
The details of the MD and TDDFT calculations, and the construction of the exciton Hamiltonian can be found in the Methods section.
Note that although classical MD provides an atomistic description of the nuclear motion, it neglects the quantum mechanical nature of high-frequency nuclear modes, thus the time dependence of the Frenkel exciton Hamiltonian in this work should be considered approximate.

\subsection{Mapping to qubits} \label{sec:encoding}
To simulate exciton dynamics with a digital quantum computer, we need to map the exciton Hamiltonian in Eq.~(\ref{eqn:frenkel}) onto qubits.
For simplicity, we use a standard binary encoding scheme in which the quantum states of an $N$-site excitonic model can be encoded in the quantum states of  $L=\log_2(N)$ qubits. 
The binary encoding method has previously been used to encode bosonic degrees of freedom with qubits in the calculation of vibrational spectroscopy~\cite{McArdle2019, Sawaya2019, Sawaya2020}.
Within the binary encoding scheme, an excitonic state $\ket{m}$ is represented by 
\begin{eqnarray} \label{eqn:state_mapping}
\ket{m} = \ket{\bm{x}} = \ket{x_1} \otimes \ket{x_{2}} \otimes. . . \ket{x_{L}},
\end{eqnarray}
where where the subscript denotes the qubit number, $m = x_1 2^0 + x_2 2^1 ... +x_{L}2^{L-1}$ and $x_i$ can be $0$ or $1$.

We note that each operator, $\ket{m}\bra{n}$, on the right-hand side (RHS) of Eq.~(\ref{eqn:frenkel}) can be mapped to qubit representation using the relation in Eq. (\ref{eqn:state_mapping})
\begin{eqnarray} \label{eqn:operator_mapping}
\ket{m}\bra{n} = \ket{\bm{x}}\bra{\bm{x'}} =  \ket{x_1}\bra{x'_{1}} \otimes \ket{x_2}\bra{x'_{2}} \otimes ... \otimes\ket{x_{L}}\bra{x'_{L}}. 
\end{eqnarray}
Each single qubit operator on the RHS of Eq. (\ref{eqn:operator_mapping}) can then be expressed in terms of Pauli and identity matrices using the following identities 
\begin{eqnarray}
\ket{0}\bra{1} = \frac{1}{2}( \hat \sigma_x +  i \hat \sigma_y )&;&
\ket{1}\bra{0} = \frac{1}{2}( \hat \sigma_x -  i \hat \sigma_y ); \nonumber \\
\ket{0}\bra{0} = \frac{1}{2}( I +  \hat \sigma_z )&;& 
\ket{1}\bra{1} = \frac{1}{2}( I -  \hat \sigma_z ) .
\end{eqnarray}

It is worth noting that the resulting Hamiltonian after encoding can be complicated since it could contain many-body interactions encompassing all $L$ qubits. 
For example, a 4-site exciton Hamiltonian can be mapped into the following two-qubit Hamiltonian (up to an identity matrix)
\begin{eqnarray} \label{eq:4site_hamiltonian}
H &=& \frac{1}{4} (E_1+E_2 - E_3 -E_4) \hat \sigma_z^1 +\frac{1}{4} (E_1 - E_2 + E_3 - E_4) \hat \sigma_z^2 \nonumber \\
&&+\frac{1}{4} (E_1 - E_2 - E_3 + E_4) \hat \sigma_z^1 \hat  \sigma_z^2 +\frac{1}{2}(V_{13} + V_{24}) \hat \sigma_x^1 \nonumber \\
&&+\frac{1}{2}(V_{12} + V_{24}) \hat \sigma_x^2 +\frac{1}{2}(V_{12} - V_{34}) \hat \sigma_z^1  \hat \sigma_x^2 \nonumber \\
&&+\frac{1}{2}(V_{13} - V_{24}) \hat \sigma_x^1  \hat \sigma_z^2 +\frac{1}{2}(V_{23} + V_{14}) \hat \sigma_x^1  \hat \sigma_x^2 +\frac{1}{2}(V_{23} - V_{14}) \hat \sigma_y^1  \hat \sigma_y^2,  
\end{eqnarray}
where the superscripts of the Pauli matrices denote the qubit indices.

\subsection{Numerical Results}\label{sec:numerics}
\begin{figure} [ht!]
  \includegraphics[width=\linewidth]{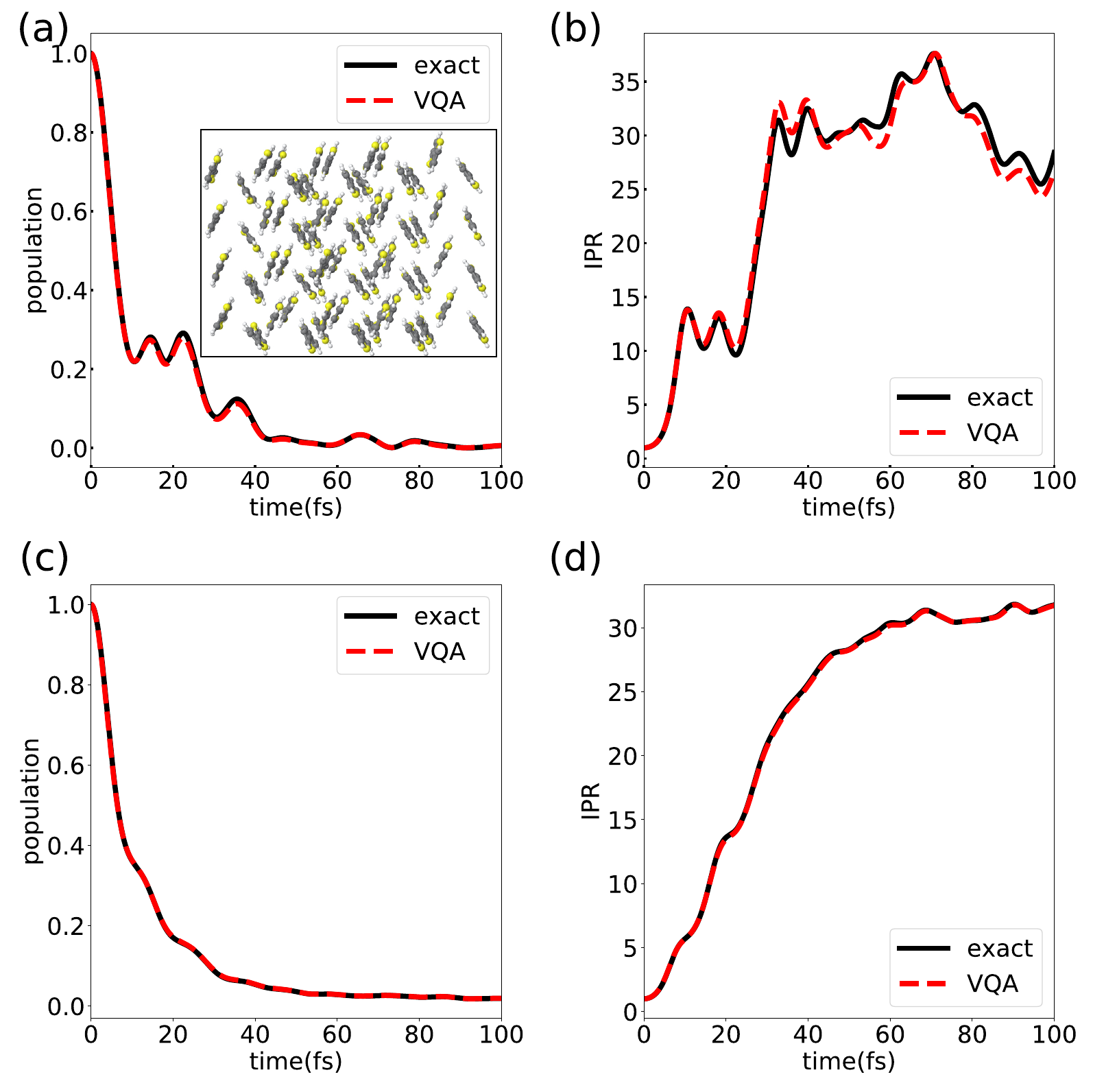}
  \caption{Frenkel exciton dynamics in a molecular crystal of 64 bi-thiophene molecules obtained from exact calculations (black solid lines) and variational quantum algorithm (red dashed lines). (a) Exciton population dynamics in molecule where the exciton is initially located.  
  (b) Time evolution of inverse participation ratio (IPR) defined in Eq.~\ref{eq:IPR}.
  (c) and (d) Dissipative exciton and IPR dynamics obtained by averaging over an ensemble of 100 pure state trajectories.}
  \label{fig:dynamics}
\end{figure}

To demonstrate the capability of VQA in modelling exciton transport, we numerically simulate the exciton dynamics in a crystal of 64 T2 molecules, arranged in a $4 \times 4 \times 2$ super cell (each unit cell contains 2 molecules). The structure of the molecular crystal is shown in the inset of Fig.~\ref{fig:dynamics}a. 
The time-dependent exciton Hamiltonian is obtained via MD and TDDFT calculations, and the details of these calculations can be found in the Methods section. 
With the binary encoding scheme described in Section \ref{sec:encoding}, the exciton Hamiltonian is encoded in 6 qubits, and a VQA simulation time-step of $0.04$fs is used.


In our simulations, a molecule at the center of the molecular crystal is initially excited, and the time evolution of the exciton wavefunction is obtained by either numerically integrating the time-dependent Schr\"{o}dinger equation, denoted as ``exact", or the VQA algorithm as described in Section \ref{sec:vqa}. Such simulation generates a single trajectory of the exciton wavefunction, and ensemble properties can be computed by averaging over many trajectories with different realizations of the Frenkel exciton Hamiltonian, sampled from MD simulations. The exciton population on the initially excited molecule computed from a single wavefunction trajectory is shown in Fig. \ref{fig:dynamics}a, and it is clear that the dynamics from VQA is in excellent agreement with that of exact calculation, confirming the capability of VQA in modelling exciton dynamics. 
From the population dynamics in Fig.~\ref{fig:dynamics}a, we observe that the electronic excitation delocalizes rapidly into neighboring molecules, the entire process takes less than 100 fs. Despite the fast delocalization, clear oscillatory behavior due to quantum coherence is also seen within the first 40 fs.

We next investigate the inverse participation ratio (IPR) of the exciton wavefunction. IPR is a global quantity that measures the extent of wavefunction delocalization: 
\begin{eqnarray} \label{eq:IPR}
    \text{IPR} = \frac{1}{\sum_m p_m^2},  
\end{eqnarray}
where $p_m$ is the probability of locating the exciton in molecule $m$. 
If the exciton is evenly delocalized across $N$ molecules, $\text{IPR}=N$ since $p_m=1/N$. On the other hand, if an exciton is fully localized in one molecule then $\text{IPR}=1$. 
Since $p_m$ is equivalent to the probability of obtaining the corresponding qubit quantum state from the output of the quantum circuit (see Eq. (\ref{eqn:state_mapping})), 
IPR is readily obtained from the distribution of the output states from a quantum computer.
In Fig.~\ref{fig:dynamics}b, the IPR numerically computed with VQA is compared to exact diagonalization results, and we again observe good agreement, further demonstrating the ability of VQA in simulating the exciton dynamics. From Fig.~\ref{fig:dynamics}b, it can be seen that IPR starts to plateau after around 40 fs due to the finite system size. Note that the maximum IPR in Fig.~\ref{fig:dynamics}b is approximately 35, which is less than the number of molecules in the system because the disorder in inter-molecular couplings as well as dynamical fluctuations cause the exciton wavefunction  unevenly distributed across all molecules, leading to a smaller IPR than the number of molecules.

We also study the dissipative exciton dynamics due to the interaction with phonon environment by averaging an ensemble of 100 pure state trajectories, the corresponding population and IPR dynamics are shown in Fig.~\ref{fig:dynamics}c and d, respectively.
We again observe nearly perfect agreement between the results from VQA and exact calculations, validating the capability of VQA in simulating exciton dynamics in complex environment. 
From Fig.~\ref{fig:dynamics}c and d, it is seen that the oscillatory behaviors of the population and IPR dynamics have disappeared due to decoherence caused by exciton-phonon interactions. Decoherence is believed to be an important driving force for efficient energy transport in photosynthetic light-harvesting systems~\cite{Mohseni2008, Rebentrost2009}.

Finally, it is worth noting that despite the exponential reduction in memory requirement due to the compact binary encoding, the measurement cost to simulate systems described by the Frenkel Hamiltonian is still proportional to the number of sites, $N$, as there can be $O(N)$ Pauli strings in the Hamiltonian. Additionally, the measurement cost to extract the population dynamics can also be significant. On the other hand, the computational advantage of quantum computers is most obvious when one goes beyond the single excitation approximation used by the Frenkel Hamiltonian.

\section{Full Hilbert Space (Multi-Exciton) Hamiltonian} \label{sec:full_hilbert}

Although the Frenkel exciton Hamiltonian is widely adopted to model energy transfer in chemical and biological systems, it restricts electronic excitation to the single-exciton manifold, neglecting the possible contribution from multi-exciton states. However, when the single-exciton states mix with the ground state\cite{Levine2006} or multi-exciton states,\cite{Spano1991,Renger1997,Bruggemann2001,May2014, Dostal2018} the Hamiltonian expressed in the full Hilbert space is more appropriate, which leads to an exponential scaling of the computational complexity with respect to the number of chromophores. Due to the prohibitive computational cost, this mulit-exciton regime has not been much explored theoretically. In this regime, a proper quantum algorithm for quantum computers may offer exponential speedup over classical simulation, and in this section, we describe how this can be achieved with the VQA algorithm.

\subsection{Mapping to Qubits}
Since each chromophore is considered as a two-level system with the ground state and the first excited state, it can be well represented by a qubit. The electronic Hamiltonian in the full Hilbert space of $N$ two-level chromophores is given by\cite{Parrish2019b}
\begin{eqnarray} 
\hat{H}(t) = \mathcal{E}(t) \hat I + \sum_{m=1}^N [ \mathcal{Z}_m(t) \hat \sigma_{z,m} + \mathcal{X}_m(t) \hat \sigma_{x,m} ] + \sum_{m=1}^N \sum_{n<m} [\mathcal{XX}_{mn}(t) \hat \sigma_{x,m} \otimes \hat \sigma_{x,n} + \nonumber \\ 
\mathcal{XZ}_{mn}(t) \hat \sigma_{x,m} \otimes \hat \sigma_{z,n} + \mathcal{ZX}_{mn}(t) \hat \sigma_{z,m} \otimes \hat \sigma_{x,n} + \mathcal{ZZ}_{mn}(t) \hat \sigma_{z,m} \otimes \hat \sigma_{z,n} ], 
\label{eq:H_martinez}
\end{eqnarray}
where $\hat I$ is the identity operator, and $\hat \sigma_{x,m}$ and $\hat \sigma_{z,m}$ are the Pauli operators associated with the $m$th chromophore. The coefficients for the corresponding Pauli operators in the Hamiltonian, $\{\mathcal{E}(t),\mathcal{Z}_m(t),\mathcal{X}_m(t),\mathcal{XX}_{mn}(t), \mathcal{XZ}_{mn}(t), \mathcal{ZX}_{mn}(t), \mathcal{ZZ}_{mn}(t)\}$, are time dependent due to the nuclear motion. Following Ref. \citenum{Parrish2019b}, we compute these coefficients from the (TD)DFT calculations, and the details can be found in Supplementary Materials.  

\subsection{Numerical Results}

\begin{figure} [ht!]
  \includegraphics[width=\linewidth]{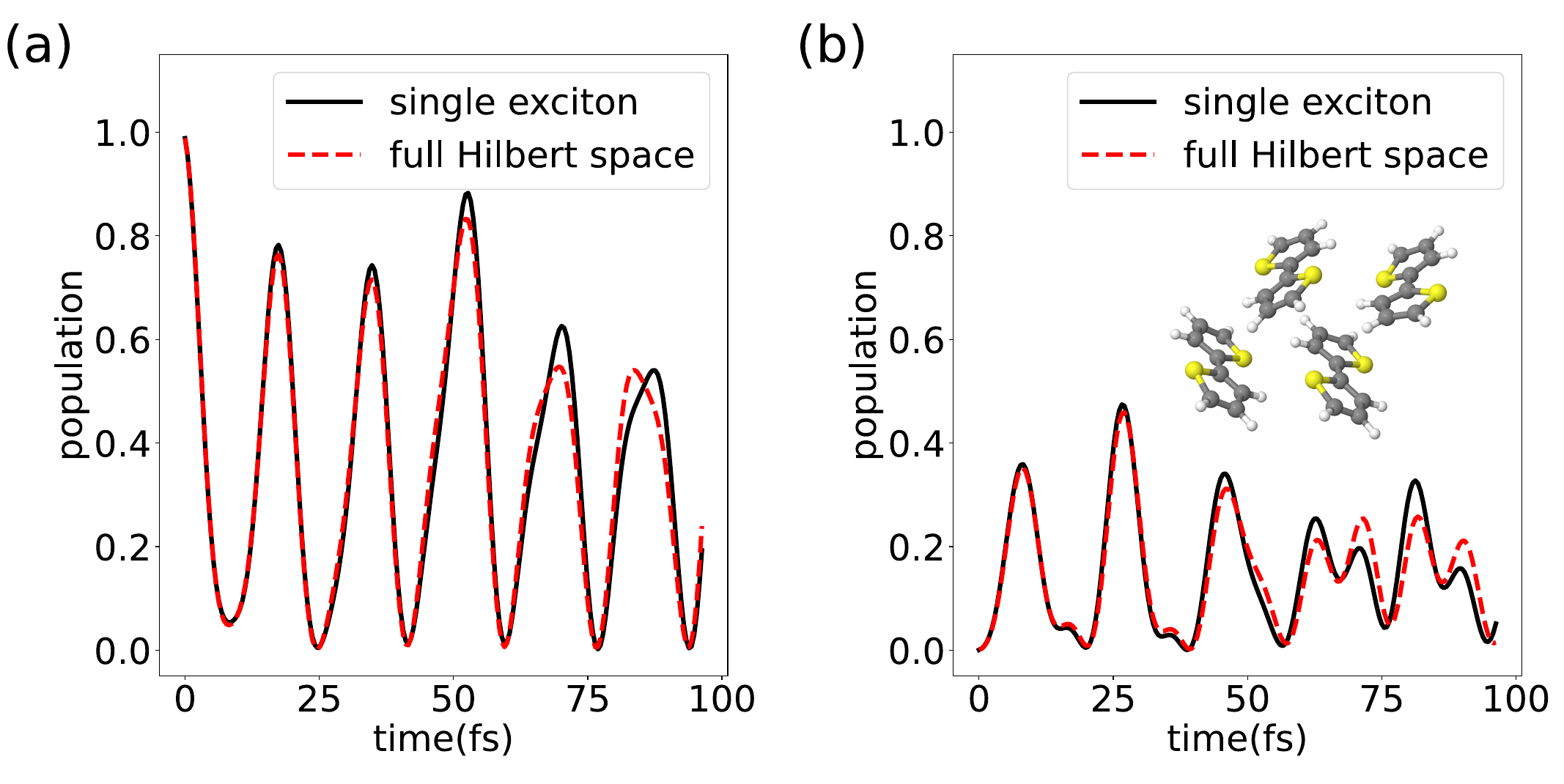}
  \caption{Full Hilbert space (red lines) and single exciton (black lines) treatments of exciton dynamics. 
  (a) Exciton population dynamics in molecule where the exciton is initially located. (b) Exciton population dynamics in initially un-excited molecule. } 
  \label{fig:fci_vs_exciton}
\end{figure}

Before showing the results from the VQA algorithm, we first compare the population dynamics computed from numerically integrating the time-dependent Schr\"{o}dinger equation governed by the Hamiltonians expressed in the Full Hibert space (Eq. (\ref{eq:H_martinez})) and in the Frenkel exciton basis (Eq. (\ref{eqn:frenkel})). For simplicity, we consider a static configuration of four T2 molecules, shown as the inset in Fig. \ref{fig:fci_vs_exciton}. Panels (a) and (b) of Fig. \ref{fig:fci_vs_exciton} show the populations of the excitation on the initially excited T2 molecule and one initially un-excited T2 molecule, respectively. Within the first 50 fs, the population dynamics from the two Hamiltonians are nearly identical, while sizable differences are seen beyond 50 fs, indicating the involvement of the multi-exciton states. It is worth noting that excitations in our T2 systems do not have strong multi-exciton character; hence Frenkel exciton Hamiltonian is a pretty good model for energy transfer. However, there are systems where state-mixing between different exciton manifolds is more substantial, such as some aggregates of light-harvesting chromophores,\cite{Parrish2019b} and full Hilbert space treatment would be more critical. 

\begin{figure} [ht!]
  \includegraphics[width=\linewidth]{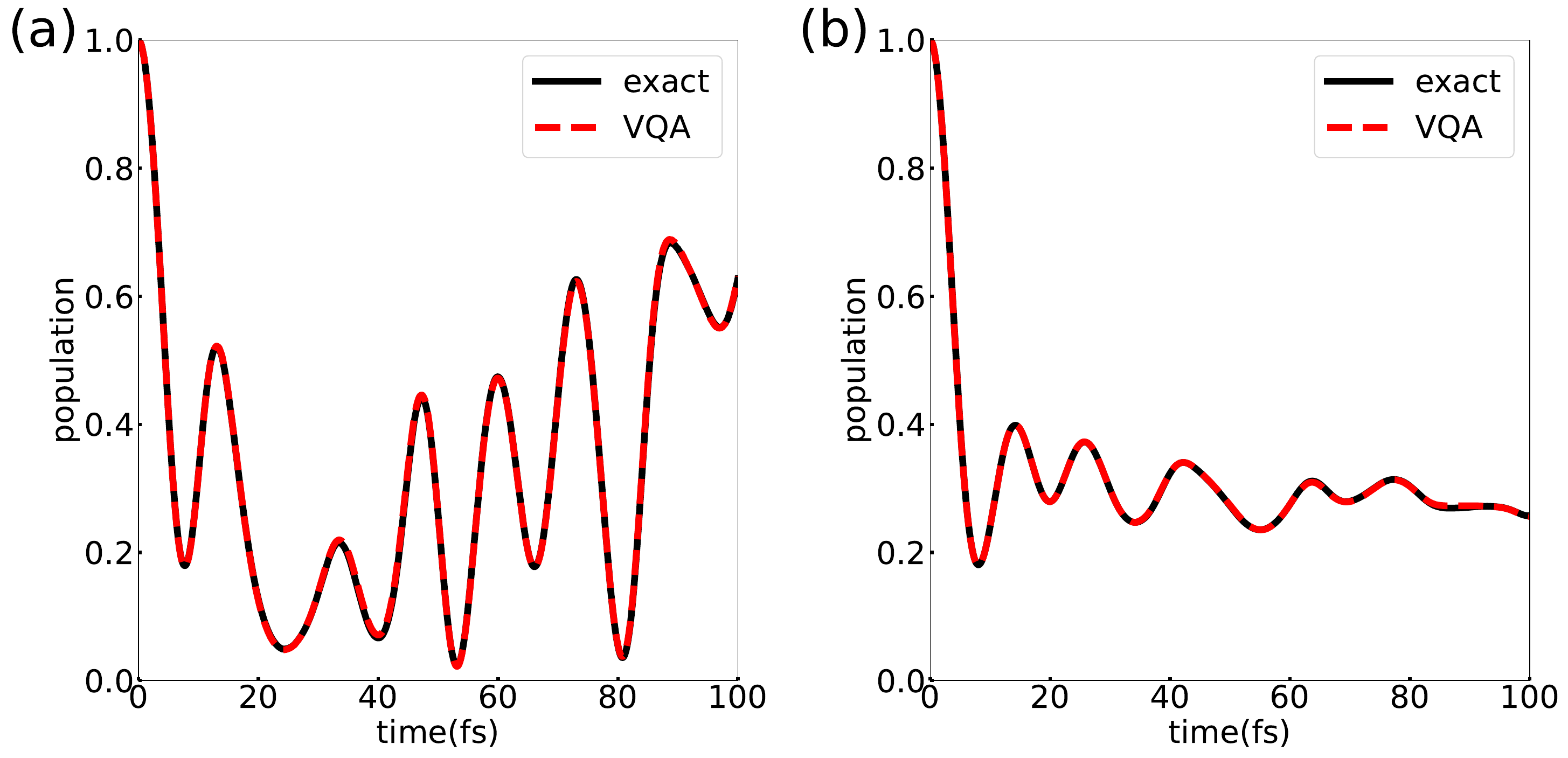}
  \caption{ Full Hilbert space treatment of exciton dynamics in a system of 4 bi-thiophene molecules, the exact calculations are shown as black solid lines and variational quantum algorithm results are displayed as red dashed lines. (a) Exciton population dynamics in molecule where the exciton is initially located.  
  (b) Dissipative exciton dynamics obtained by averaging over an ensemble of 100 pure state trajectories.}
  \label{fig:fci_dynmaics}
\end{figure}

We now consider the capability of the VQA algorithm in reproducing the ``exact" population dynamics, which is obtained by numerically integrating the time-dependent Schr\"{o}dinger equation with the Hamiltonian in the full Hilbert space. Fig. \ref{fig:fci_dynmaics}(a) shows the excitation population on the initially excited T2 molecule from a single trajectory of a T2 tetramer whose time-dependent Hamiltonian is obtained from MD simulation and quantum-chemical calculations. Almost perfect agreement is seen between the VQA and exact results. Such excellent agreement remains when we calculate the average population dynamics over an ensemble of 100 trajectories, as shown in Fig. \ref{fig:fci_dynmaics}(b). This shows the promise of using quantum computers to simulate energy transfer for systems that have significant state mixing. 

\section{Experimental Results on IBM Quantum Computers}\label{sec:experiment}
\begin{figure} [ht!]
  \includegraphics[width=.9\textwidth]{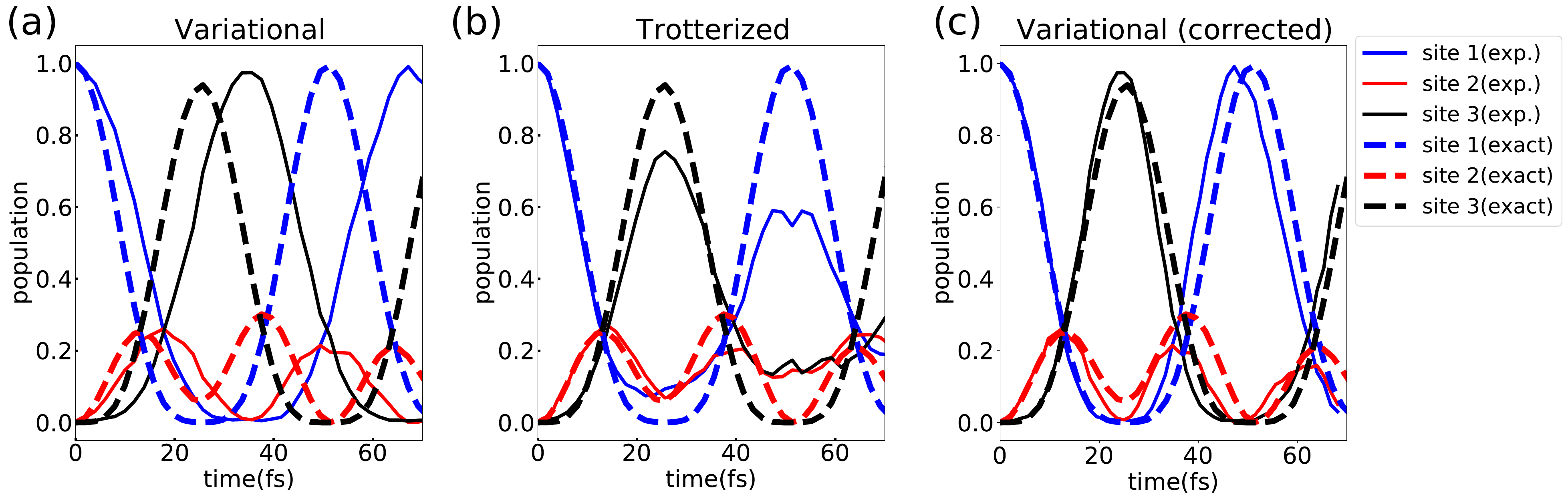}
  \caption{Frenkel exciton population dynamics in a linear chain of 4 bi-thiophene molecules (encoded in 2 qubits) obtained from IBM quantum computer (solid lines) and from exact calculations (dashed lines). 
  The experimental results are obtained using (a) variational quantum algorithm, (b) Trotter scheme and (c) variational quantum algorithm with error mitigation.}
  \label{fig:ibm_dynamics}
\end{figure}

We assess the capability of current digital quantum computers in simulating the Frenkel exciton dynamics by studying a linear chain of 4 T2 molecules with an IBM quantum computer.
Due to the limitations of the actual quantum device, here, we consider a simplified model and assume a static system with identical nearest neighbor coupling of $V=40$ meV and periodic boundary condition. This value of coupling corresponds to 2 T2 molecules separated by approximately 5.6\AA\; and the plane of molecules are arranged in parallel to each other.  Furthermore, we assume site energies of  $E_1=E_2$, $E_3=E_4$ and $E_1 - E_3 = \Delta E = 20$meV. 
With these approximations, the 4-molecule system is encoded in 2 qubits and the resulting Hamiltonian is $H = \frac{\Delta E}{2} \sigma_z^1 + V  \sigma_x^2 + V  \sigma_x^1  \sigma_x^2$.
We use a wavefunction ansatz with only three variational parameters $\ket{\psi(\vec{\theta})} = \mbox{e}^{i \theta_3 \sigma^1_x \sigma^2_x}  \mbox{e}^{i \theta_2 \sigma^2_x}  \mbox{e}^{i \theta_1 \sigma^1_z} \ket{\psi_0}$, corresponding to the popular Hamiltonian ansatz~\cite{Wecker2015}.
The compiled quantum circuits used in obtaining $\hat M$ and $\vec V$ in Eq.~(\ref{eq:EOM}) can be found in Supplementary Materials (SM). 
Molecule 1 is initially excited, and a time step of $\delta t = 1.97$fs is used (i.e. 3$\hbar \mbox{eV}^{-1}$) throughout the simulations in this section. 
The quantum computer used here is the 5-qubit system code-named $ibmq\_rome$. 

Fig.~\ref{fig:ibm_dynamics}a displays the population dynamics computed by IBM quantum computer using the VQA algorithm (solid lines). For comparison, we also include the exciton dynamics computed exactly by numerically integrating the Schr\"{o}dinger equation (dashed lines). 
For clarity, we only show the population dynamics of site 1 to site 3 since the population dynamics of site 4 is similar to that of site 2. The dynamics of site 4 population is found in SM. 
From Fig.~\ref{fig:ibm_dynamics}a, it can be seen that though VQA correctly captures the amplitudes of the oscillations of different sites, the oscillation frequencies from VQA are smaller than that of the exact dynamics and this results in an overall right shift of the VQA dynamics.
The accuracy of the VQA dynamics is principally determined by three factors: the expressive power of the wavefunction ansatz, error due to finite number of measurements and the imperfections of quantum devices.  To identify the main source of error, we performed additional calculations (see SM) with noiseless quantum simulator and show that the discrepancy with exact results observed in Fig.~\ref{fig:ibm_dynamics}a is largely due to the imperfection of the quantum devices. We observe that the noiseless simulator results using the same number of measurements are in good agreement with exact calculations. 

It is also instructive to study the exciton dynamics with the standard Trotterization method by discretizing the evolution operator: $\ket{\Psi(t)} = \prod_n \mbox{e}^{-i H \delta t} \ket{\Psi(0)}$ where $\delta t = t/n$. 
For the 4-molecule system considered here, each discrete operator $\mbox{e}^{-i H \delta t} \approx \mbox{e}^{-i V\sigma_x^1 \sigma_x^2 \delta t} \mbox{e}^{-i V\sigma_x^1  \delta t} \mbox{e}^{-i \Delta E \sigma_z^1  \delta t}$ can then easily be implemented with single and two-qubit rotations. 
The Trotter results from IBM quantum computer are shown in Fig.~\ref{fig:ibm_dynamics}b (dashed lines).
As opposed to VQA, the results from the Trotterization scheme fail to capture the amplitudes of the oscillations, and the decreasing amplitude indicates that decoherence effect is significant. 
On the other hand, from Fig.~\ref{fig:ibm_dynamics}b, it is seen that the Trotterization approach captures the periods of the exciton dynamics since the peak and trough positions coincide with those from the numerically exact results.

\begin{figure}[ht!]
  \includegraphics[width=4.5in]{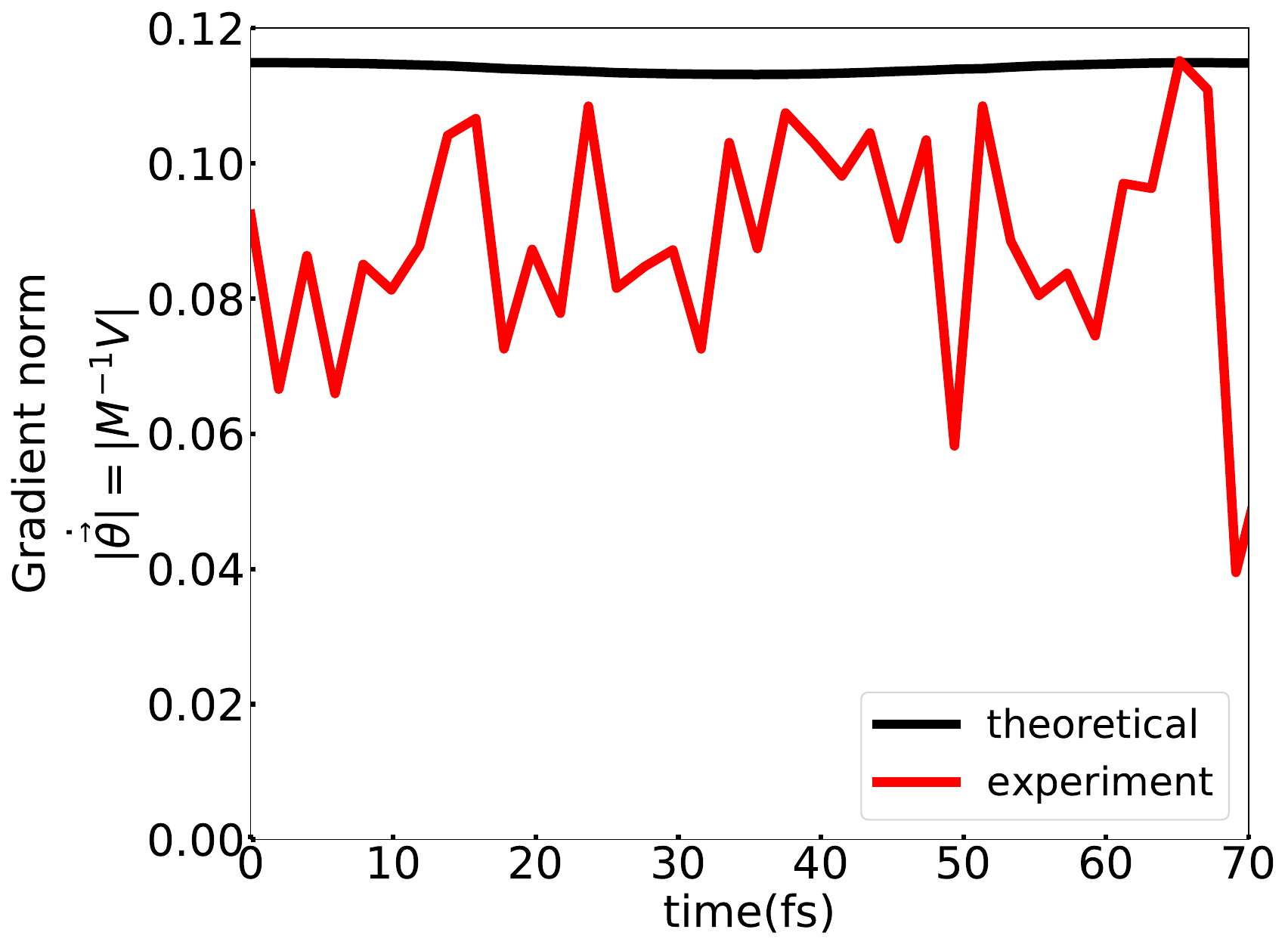}
    \caption{The norms of gradient, $\dot{\vec{\theta}}$, obtained from IBM quantum computer (solid red line) and the theoretical values (solid black lines). The theoretical value is almost always larger than that of the experimental value due to errors in quantum hardware.  }
    \label{fig:gradient}
\end{figure}

\section{Error Analysis and Mitigation} \label{sec:mitigation}
To identify the origin of the discrepancy between the VQA and exact dynamics in Fig.~\ref{fig:ibm_dynamics}a, we investigate the norm of the gradient in Eq.~(\ref{eq:EOM_theta}), $\dot{\vec{\theta}} = \hat M^{-1} \vec V$, and the results are shown in Fig.~\ref{fig:gradient}. 
The red line denotes the magnitude of $|\dot{\vec{\theta}}|$ obtained from the IBM quantum computer whereas the black line is the theoretical value for the same set of variational parameters.
It is found that the gradient norms from the quantum computer are almost always lower than the theoretical ones due to the errors in the quantum hardware.

\begin{figure}[ht!]
  \includegraphics[width=4.5in]{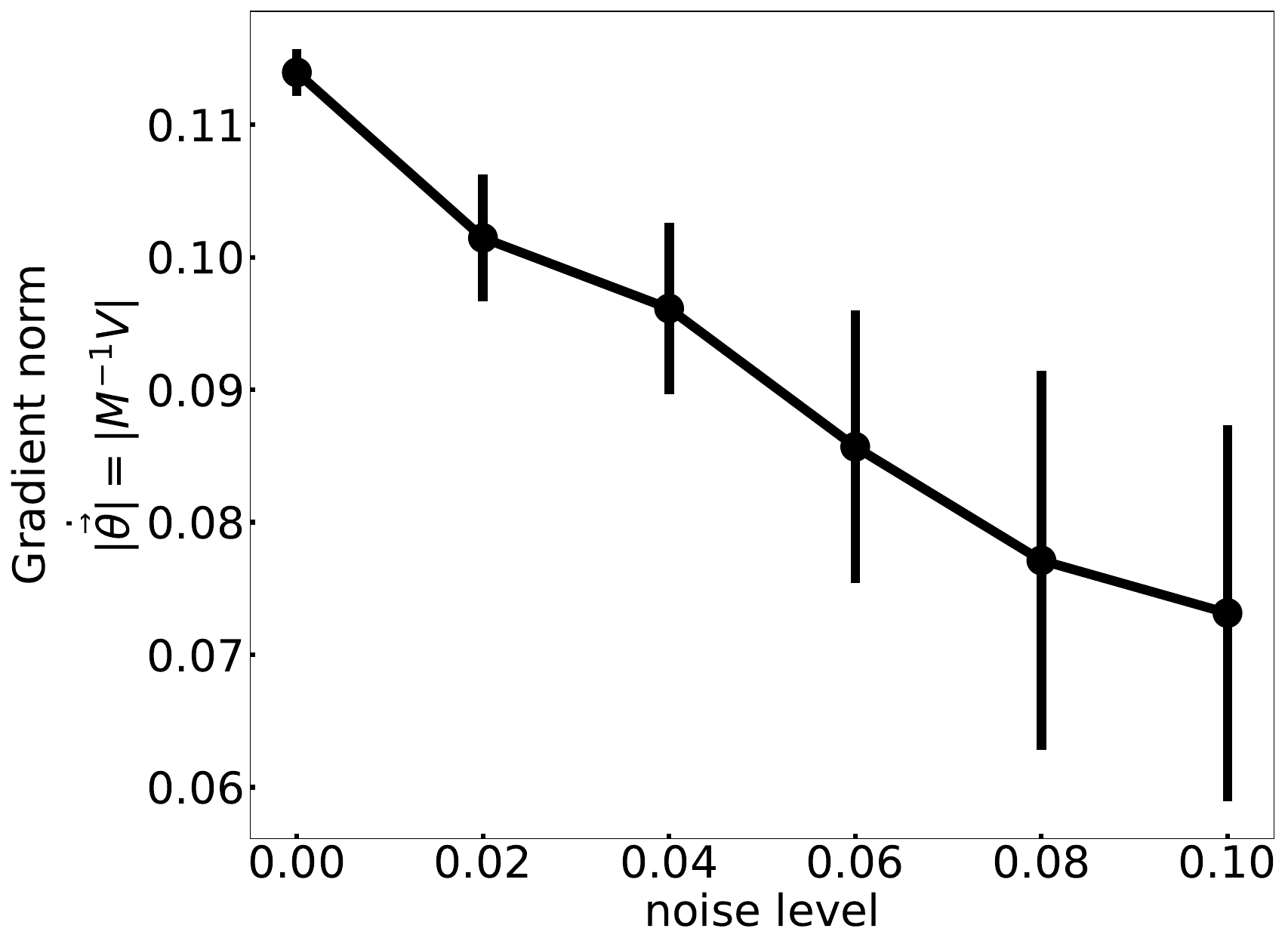}
    \caption{Dependence of the gradient norm, $|\dot{\vec{\theta}}|$, as a function of noise level using an artificial noise model in quantum simulator. The error bars denotes the standard deviations from 100 independent simulations. The error bar at zero noise arises from the finite number of measurements.}
    \label{fig:gradient_noise_model}
\end{figure}

To gain insights into the underestimation of the gradient, it will be useful to analyse the effects of small perturbation on $\dot{\theta}$. Following Ref.~\cite{Endo2019}, we denote $\hat M = \hat M_0 + \delta \hat M$, $\vec V=\vec V_0 + \delta \vec V$ and $\dot{\vec \theta} = \dot{\vec \theta}_0 + \delta \dot{\vec \theta}$, where the subscript $0$ represents the exact quantities without noise, and preceding $\delta$ represents the corrections due to the hardware or shot noise, thus
\begin{eqnarray}
   \delta \dot{ \vec \theta} &=& \hat M^{-1} \vec V - \hat M_0^{-1} \vec V_0, \nonumber \\
                       &=& (\hat M_0 + \delta \hat M )^{-1}( \vec V_0 + \delta \vec V) - \hat M_0^{-1} \vec V_0,  \nonumber \\
                       &\approx& \hat M_0^{-1} \delta \vec V -  \hat M_0^{-1}\delta \hat M \hat M_0^{-1} \vec V_0, \label{eq:theta_error}
\end{eqnarray}
where we use Taylor expansion and only keep the first order terms to obtain the last line.

We assume a depolarizing noise model, i.e.
\begin{eqnarray}
    E_{\text{depolarizing}}(\hat \rho) = (1-\lambda) \hat \rho + \lambda \hat I/2^N,
\end{eqnarray}
where $N$ is the number of qubits, $\hat \rho$ is the density matrix and $\lambda$ is the noise level. The expectation value of an operator, $A$, then reduces to $ \langle A \rangle= (1- \lambda) \langle A \rangle_0 $ where the subscript again denotes the noiseless value. Note that we make use of relation $\text{Tr}[A] = 0$ which is valid for Pauli strings. 
Applying the depolarising model to Eq.~(\ref{eq:theta_error}), we then have 
\begin{eqnarray}
   \delta \dot{\theta} = (\lambda_M - \lambda_V) M_0^{-1} V_0, \label{eq:theta_erro}
\end{eqnarray}
where $\delta M = - \lambda_M M_0$ and $\delta V = - \lambda_V V_0$, assuming all the matrix elements in $V$ or $M$ are equally affected by noise. Given that the number of quantum operations in obtaining $M$ is less than that of $V$ due to partial cancellation of unitary operators with the corresponding transposed operators (see Eq.~(\ref{eqn:circuit})), it is expected $\lambda_M < \lambda_V$, leading to underestimation of the magnitude of $|\dot{\theta}|$.

We next numerically investigate how the norm of $\dot{\vec{\theta}}$ depends on noise level in quantum devices by performing noisy simulations using the IBM Qiskit quantum simulator~\cite{Qiskit}. The details of the noise model can be found in SM. 
In Fig.~(\ref{fig:gradient_noise_model}), we compute the norm of gradient as a function of noise level, $\lambda$, using the variational parameters obtained experimentally at $t=40$fs. We use 8192 shots in the simulations. It can be seen that the norm $|\dot{\vec{\theta}}|$ decreases monotonically as a function of $\lambda$, an observation consistent with the experimental observations that noise in quantum devices leads to an under-estimation of gradient norm. 
The under-estimation of the gradient seen in Fig.~\ref{fig:gradient} therefore leads to an effective update time-step that is smaller than the actual time-step, i.e. $\delta t_{\mathrm{eff}} < \delta t$, leading to the right-shift of the VQA dynamics compared to the exact results. 

Next we show that this error can be partially mitigated by replacing $\delta t$ with $\delta t_{\mathrm{eff}}$ in the equation of motion of the variational parameters, Eq.~\ref{eq:EOM_theta}. 
Writing $\delta t_{\mathrm{eff}} = \delta t/\alpha$ where the correction factor $\alpha > 1.0$, we propose extracting the optimal $\alpha$ from the short-time dynamics of the Trotter simulations. 
While the Trotter scheme is not suitable for NISQ devices due to the need for deep circuit depth, the short-time dynamics is expected to be more accurate than VQA since only small number of quantum gates are executed, and there is no need for the complicated Hadamard tests in constructing $\hat M$ and $ \vec V$. Specifically, we choose the optimal $\alpha$ by minimizing the difference between site 1 populations obtained from VQA and from the Trotter scheme, i.e.
\begin{eqnarray}
   \min_\alpha \int_0^{t_c} \Big (p^{\text{VQA}}_{1}(t) - p^{\text{Trotter}}_{1}(t) \Big)^2dt,
\end{eqnarray}
where we use a cut-off time of $t_c = 20$fs. In our case, we found the optimal $\alpha$ to be 1.42, a value close to the ratio between the time-averaged theoretical and experimental norms shown in Fig.~\ref{fig:gradient} (i.e. 1.34). The corrected VQA dynamics are shown in Fig.~\ref{fig:ibm_dynamics}c, it can be seen that the corrected results accurately reproduce the population evolution of all the sites in the system though only site 1 population is used to extract the optimal $\alpha$. Thus by combining the results from both algorithms, the error-mitigated exciton dynamics can be significantly more accurate than the results from either VQA or Trotter scheme.

In SM, we demonstrate the above error mitigation method in a different system by simulating the dynamics of transverse field Ising (TFI) model with noisy simulator. We again obtained a much improved results compared to the simulations without error mitigation, suggesting the generality of our approach.
The error mitigation technique above can be generalized to more complex systems in which the correction factor, $\alpha$, may not be a simple constant but could vary with time. In such cases, we could divide the total simulation time into multiple simulation windows. For each simulation window, we perform a separate Trotter simulation to extract the optimal shift factor for that particular segment, with initial condition being the variational quantum state at the beginning of the time window. 
Finally, it is worth noting that our proposed error mitigation method can be easily combined with other error mitigation methods, such as those based on noise extrapolation, to maximize the performance of quantum computers. 
\section{Discussions and Conclusions}
Most of the recent experimental and theoretical NISQ research for chemical applications focus on finding the ground or low energy states of molecular systems. Here we expand the scope of applications for near-term digital quantum computers to time-dependent processes in many-body chemical systems.
Compared to the ground state calculations, simulating quantum dynamics is typically a more challenging task as it often involves the participation of all quantum states.
NISQ algorithms for ground state simulation like the popular variational quantum eigensolver (VQE) involve minimizing the expectation value of the Hamiltonian for a given variational wavefunction, and the optimization path is usually not important as long as the final optimized wavefunction is a good approximation of the true ground state wavefunction of the Hamiltonian. 
On the other hand, VQA for quantum dynamics requires the path of the variational wavefunction to closely match that of the exact dynamics throughout the entire evolution, and a small discrepancy can easily lead to a very different quantum state at a later time.

Studying energy transfer in molecular systems is a fitting example to test the capability of quantum computers in simulating quantum dynamics in chemical systems. 
Understanding energy transfer process in molecular systems is important with a wide range of practical applications, e.g. design of efficient OPVs and OLEDs. 
However, accurate simulation of exciton dynamics for realistic systems is challenging in classical computers. 
The presence of disorder and long-range couplings in molecular systems renders Bloch's band theory commonly used in condensed matter physics inapplicable due to the lack of symmetry.
Coarse-grained or classical models can provide important physical insights, but fail to capture the quantum mechanical properties and chemical details of the actual materials. Quantum computers could therefore potentially overcome these computational challenges.
With the binary encoding scheme, a Frenkel exciton Hamiltonian of $N$ sites can be encoded in $\log_2(N)$ qubits.
In other words, three dimensional molecular systems consisting of millions of sub-units can be encoded in as few as tens of qubits, albeit the dynamical simulation still requires significant measurement cost that could scale polynomially with the number of sites.
The advantage of quantum computers is most obvious for simulating the dynamics of multi-exciton Hamiltonian, since both the circuit depth and measurement cost scale polynomially with system sizes whereas the classical computational cost scales exponentially. For example, simulating the multi-exciton dynamics of 50 to 60 sites is a formidable task even for the most powerful classical computers. On the hand, quantum computers with more than 60 qubits already exist in the super-conducting platform~\cite{ibm_roadmap, Wu2021}.

While our numerical calculations affirms the capability of the VQA in simulating  quantum dynamics of molecular systems, running the algorithm on actual quantum computer provides us with important insights into the robustness of the algorithm against device imperfections. 
While experimental demonstration of hybrid algorithms for ground-state quantum chemistry has been performed on systems that exceed 10 qubits~\cite{AIQuantum2020}, digital quantum simulations of time-dependent chemistry problems has been lacking.
In fact, the time-dependent VQA has only been recently demonstrated on a 2-qubit Ising model in the adiabatic limit~\cite{Chen2020}.
Indeed our experimental results with IBM quantum devices demonstrate that, despite only short circuit depth is needed in VQA, the error rates in current quantum computers are still too large to accurately simulate the quantum dynamics of molecular Hamiltonians with a few qubits.

Despite the failure of VQA (without error mitigation) in capturing the correct quantum dynamics, analysis of the results provides important insights into the source of errors and how to mitigate them.
Our investigation reveals that the deviation from the exact result arises from the under-estimation of the gradients, $\dot{\vec{\theta}}$ as shown in Fig.~\ref{fig:gradient}. 
We further show that this error in gradient estimation can be largely remedied by extracting a correction factor from the short-time dynamics of the Trotter simulation.
To the best of our knowledge, our work constitutes the first demonstration that energy transport in realistic chemical systems can be accurately simulated in a digital quantum computer.  

Recently there have been several experimental demonstration of error mitigation techniques for ground-state quantum chemistry problems with encouraging results~\cite{Kandala2019, AIQuantum2020}, our work provides an additional tool to combat quantum device errors for dynamical simulations in chemistry.
The effectiveness of the error mitigation method shown in Fig.~\ref{fig:ibm_dynamics}c also opens the door to the possibility of improving the accuracy of quantum simulations by combining the experimental results from different quantum algorithms. 
Unlike many other error mitigation methods that require multiple executions of the quantum circuits to construct the statistical profile of the errors, our error mitigation method only requires running each quantum algorithm once. Thus this is particularly appealing for cloud based NISQ devices in which the queue time and cost could be significant.
This error mitigation method can be extended to simulate other time-dependent phenomena in biology and chemistry such as spectroscopy, chemical reaction and charge transfer.
A systematic approach to perform this error mitigation technique in combination with other error mitigation methods will be a fruitful future endeavor, though it is beyond the scope of this current work. 

To conclude, in this work, we propose the simulation of energy transfer dynamics in molecular systems with digital quantum computers. We first perform multi-scale calculations to obtain the exciton Hamiltonian. The Hamiltonian is then mapped onto qubits, and we use a hybrid time-dependent VQA to simulate the time evolution of the encoded Hamiltonian. Our numerical example affirms the feasibility of this approach.
Despite the short circuit depth in VQA, our experimental results with IBM quantum computer show that current quantum devices are still plagued with too much noise for a good simulation of the dynamics of these quantum many-body systems.
We propose a new error mitigation technique for correcting the VQA dynamics with the information extracted from the short-time dynamics of Trotter simulation. We demonstrate that the corrected VQA dynamics is significantly improved and  capable of capturing the true dynamics of the excitonic system. 
Our work extends the scope of applications of NISQ devices to time-dependent processes in chemical systems, and opens the door to new error mitigation technique that combines the experimental results of multiple quantum algorithms. 

\section{Methods}
To construct the time-dependent Frenkel exciton Hamiltonian, classical MD simulation was performed for a T2 crystal with 64 T2 molecules in the simulation box. The initial simulation box was set up as a $4\times4\times2$ super cell based on the experimental~\cite{Pelletier1994} crystal structure of T2 (CSD identifier: DTENYL02). OPLS/2005 force field was employed in the simulation as it can reasonably reproduce the torsional potential energy surface of T2 predicted from the localized second order M{\o}ller-Plesset perturbation theory (LMP2)~\cite{Dubay2012}. The MD simulation was performed with the Desmond package 3.6~\cite{Bowers2006} in the NVT ensemble at 133 K, the crystal temperature in the experiment~\cite{Pelletier1994}. Temperature was maintained by the Nos\'{e}-Hoover thermostat with a coupling constant of 2.0 ps. Periodic boundary condition was applied to the monoclinic simulation box, and the particle-mesh Ewald (PME) method was employed for electrostatic interactions. The simulation time step was 1 fs, and the configurations were saved every 2 fs during the 10-ps production run. A snapshot of the MD configuration is given in Fig.~\ref{fig:dynamics}a. 

The site energy in the Frenkel exciton Hamiltonian is approximated by the lowest-lying excited-state energy of each T2 in the system, which was computed by TDDFT with the Tamm-Dancoff approximation (TDA) using CAM-B3LYP/6-31+G(d). In our previous work~\cite{Lu2020} we have shown that the excited-state energy of T2 predicted by CAM-B3LYP/6-31+G(d) agrees well with that from a correlated wavefunction method (CC2 method). All the quantum chemical calculations were performed with the PySCF program~\cite{Sun2018}, and density fitting was used with the heavy-aug-cc-pvdz-jkfit auxiliary basis set implemented in PySCF. In estimating the excitonic coupling between the local excitations on two T2 molecules ($m$ and $n$), $V_{mn}$, we computed the Coulomb coupling via\cite{Farahvash2020}
\begin{equation}
    V_{mn} \approx 2 \sum_{iajb} X_{ia}^{(m)} X_{jb}^{(n)} \iint  \psi_i^{(m)}(\vec{r}_1) \psi_a^{(m)}(\vec{r}_1) \frac{1}{|\vec{r}_1-\vec{r}_2|}\psi_j^{(n)}(\vec{r}_2) \psi_b^{(n)}(\vec{r}_2) d\vec{r}_1 d\vec{r}_2,
\label{eq:coupling}
\end{equation}
where $X_{ia}^{(m)}$ is the excitation amplitude for the electronic transition from the occupied molecular orbital (MO), $\psi_i^{(m)}$, to the virtual MO, $\psi_a^{(m)}$, of molecule $m$. In using the Coulomb coupling to approximate the excitonic coupling, we have neglected the contributions from the Hartree-Fock exchange and the exchange-correlation of the employed functional, CAM-B3LYP. The computations of site energies and couplings were performed for the 5,000 frames harvested from the MD simulation, leading to a trajectory of time-dependent Frenkel exciton Hamiltonian. 

For the full Hilbert space Hamiltonian (Eq. (\ref{eq:H_martinez})), we consider a static configuration of four T2 molecules extracted from the crystal of 64 T2 molecules, shown as the inset in Fig. \ref{fig:fci_vs_exciton}. For the coefficients in Eq. (\ref{eq:H_martinez}), we followed Ref. \citenum{Parrish2019b}, and obtained them from the DFT and TDDFT calculations. The calculation details are provided in the SM.  

It is worth noting that even though the computational cost of constructing the exciton Hamiltonian through quantum chemistry calculations scales linearly with the number of sub-units (assuming that the inter-molecular couplings are negligible beyond certain distance), the computational cost itself is still high for large systems. However, several statistical and machine-learning methods have been developed to overcome this difficulty~\cite{Farahvash2020, Lee2020, Lu2020}. Once sufficient quantum chemistry data have been generated, the remaining matrix elements in the exciton Hamiltonians can be obtained through these statistical or machine learning methods with very high accuracy.

\section{Conflicts of interest}
There are no conflicts to declare.

\section{Author Contribution}
CKL conceived the initial idea and developed it along with LS. CKL performed the numerical quantum simulations, JL ran the experiments on IBM Q devices and LS performed the MD and TDDFT calculations. All authors contribute to the discussions and preparation of manuscript. 

\section{Data Availability}
The data generated in this study are available from the corresponding author on reasonable request.

\section{Acknowledgment}
L.S. acknowledges the support from the University of California Merced start-up funding. J.W.Z.L and L.C.K. would like to thank the National Research Foundation and the Ministry of Education, Singapore, for financial support.

\clearpage
\bibliographystyle{naturemag_noURL}
\bibliography{MyCollection}

\end{document}


\title{Supplementary Materials: Quantum Simulations of Energy Transfer in Molecular Systems with Digital Quantum Computers}

\author{Chee-Kong Lee}
\email{cheekonglee@tencent.com}
\affiliation{Tencent America, Palo Alto, CA 94306, United States}
\author{Jonathan Wei Zhong Lau}
\email{e0032323@u.nus.edu}
\affiliation{Centre for Quantum Technologies, National University of Singapore, 117543, Singapore}
\author{Liang Shi}
\email{lshi4@ucmerced.edu}
\affiliation{Chemistry and Chemical Biology, University of California, Merced, California 95343, United States}
\author{Leong Chuan Kwek}
\email{cqtklc@nus.edu.sg}
\affiliation{Centre for Quantum Technologies, National University of Singapore, 117543, Singapore}
\affiliation{National Institute of Education, Nanyang Technological University, 1 Nanyang Walk, Singapore 637616}
\affiliation{MajuLab, CNRS-UNS-NUS-NTU International Joint Research Unit, UMI 3654, Singapore}

\maketitle
\section{Artificial Noise Model}
\label{sec:noise_model}

In Fig. 6 of the main text we investigate the dependence of gradient, $\dot{\vec{\theta}} = \hat M^{-1} \vec V$, on noise level using an artificial noise model with IBM Qiskit quantum simulator~\cite{Qiskit}.
The artificial noise model was created using the tools in Qiskit Aer noise model. Specifically in the noise model, we defined an overall noise parameter $\lambda$, where $0\leq \lambda \leq 1$. This parameter controls the depolarizing error that is added to all single and two qubit rotations and control gates (e.g. CNOT gate). The map for the depolarizing channel is
\begin{gather}
    E_{\text{depolarizing}}(\rho) = (1-\lambda)\rho + \lambda \text{Tr}[\rho]I.
\end{gather}
The depolarizing error, $\lambda$, is therefore the probability that a single qubit, after undergoing a rotation, will be replaced by the completely mixed state. It is usually considered to be the quantum version of the independent bit flip error in classical devices. 


\section{Additional Results from IBM Quantum Computer}
\begin{figure}[H]
  \includegraphics[width=\linewidth]{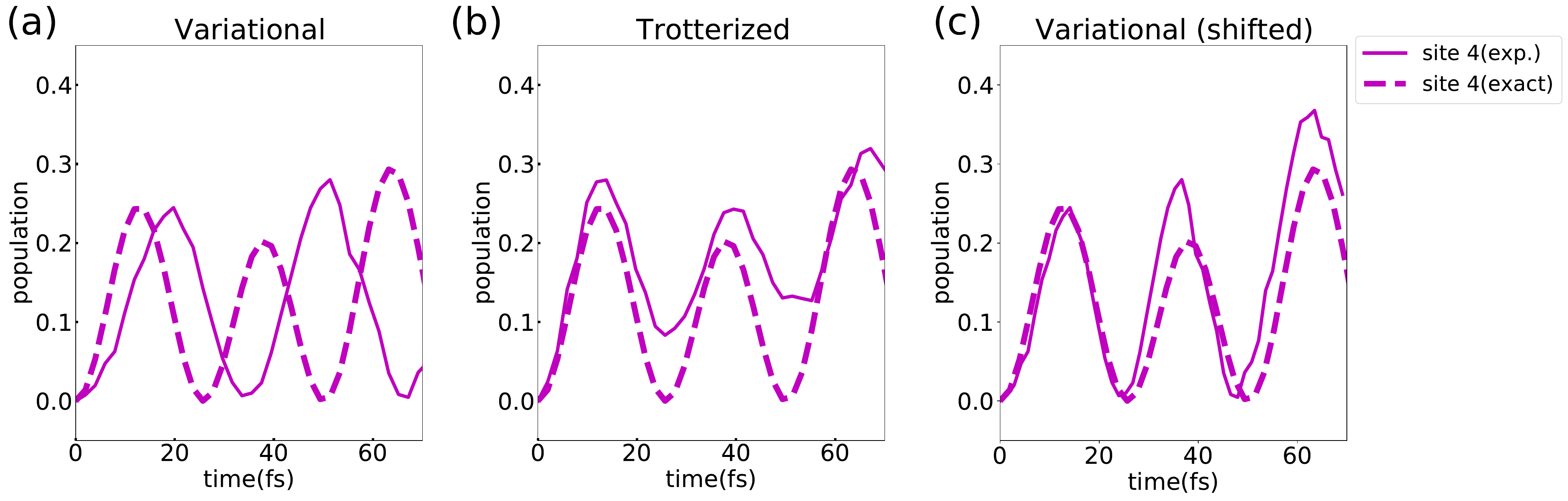}
    \caption{The dynamics of site 4 population in Sec. VI obtained from (a) variational quantum algorithm (VQA), (b) Trotter scheme and (c) corrected VQA.}
    \label{fig:ibm_site4}
\end{figure}

The dynamics of site 4 population in Sec.~VI obtained from variational quantum algorithm (VQA), Trotter scheme and corrected VQA are shown in Fig.~\ref{fig:ibm_site4}.

\section{Quantum circuits for the 4-molecule systems}
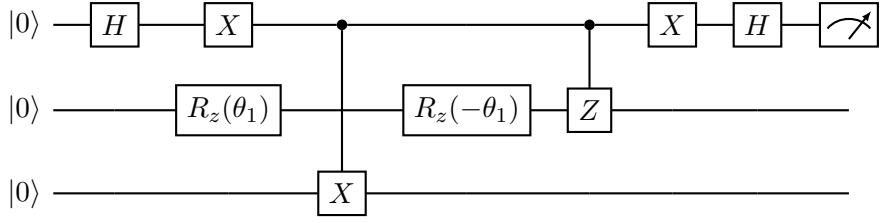
\begin{figure}[H] 
    \centering
    \begin{tikzcd}
    \lstick{$\ket{0}$} & \gate{H} & \gate{X}&\ctrl{2}&\qw&\ctrl{1}&\gate{X}&\gate{H}&\meter{}\\
    \lstick{$\ket{0}$} & \qw& \gate{R_z(\theta_1)}&\qw&\gate{R_z(-\theta_1)}&\gate{Z}&\qw&\qw&\qw\\
    \lstick{$\ket{0}$} & \qw& \qw&\gate{X}&\qw&\qw&\qw&\qw&\qw
    \end{tikzcd}
    \caption{Quantum circuit for computing $\mbox{Re} ( \left\langle  \frac{\partial \psi} {\partial \theta_1} \right\vert \left.  \frac{\partial \psi}{\partial \theta_2} \right\rangle )$.}
    \label{fig:M_12}
\end{figure}

For the 2-qubit simulations in Sec. VI in the main text, the compiled quantum circuits for computing the matrix elements of $\hat M$ are shown in Fig.~\ref{fig:M_12}, \ref{fig:M_13} and \ref{fig:M_23}. 

\begin{figure}[H] 
    \centering
    \begin{tikzcd}
    \lstick{$\ket{0}$} & \gate{H} & \gate{X}&\ctrl{2}&\ctrl{1}&\qw&\ctrl{1}&\gate{X}&\gate{H}&\meter{}\\
    \lstick{$\ket{0}$} & \qw& \gate{R_z(\theta_1)}&\qw&\gate{X}&\gate{R_z(-\theta_1)}&\gate{Z}&\qw&\qw&\qw\\
    \lstick{$\ket{0}$} & \qw& \gate{R_x(\theta_2)}&\gate{X}&\qw&\qw&\qw&\qw&\qw&\qw
    \end{tikzcd}
    \caption{Quantum circuit for computing $\mbox{Re} ( \left\langle  \frac{\partial \psi} {\partial \theta_1} \right\vert \left.  \frac{\partial \psi}{\partial \theta_3} \right\rangle )$.}
    \label{fig:M_13}
\end{figure}

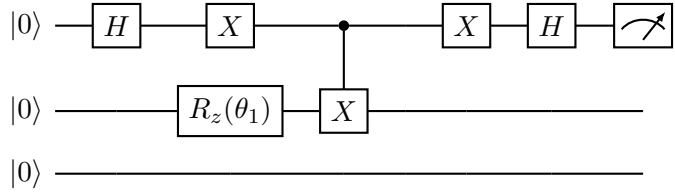
\begin{figure}[H] \label{fig:circuit_M23}
    \centering
    \begin{tikzcd}
    \lstick{$\ket{0}$} & \gate{H} & \gate{X}&\ctrl{1}&\qw&\gate{X}&\gate{H}&\meter{}\\
    \lstick{$\ket{0}$} & \qw& \gate{R_z(\theta_1)}&\gate{X}&\qw&\qw&\qw&\qw\\
    \lstick{$\ket{0}$} & \qw&\qw&\qw&\qw&\qw&\qw&\qw
    \end{tikzcd}
    \caption{Quantum circuit for computing $\mbox{Re} ( \left\langle  \frac{\partial \psi} {\partial \theta_2} \right\vert \left.  \frac{\partial \psi}{\partial \theta_3} \right\rangle )$.}
    \label{fig:M_23}
\end{figure}
\noindent where $X$, $Y$ and $Z$ denote the Pauli matrices. 

Nine quantum circuits are needed for computing $\vec V$ vector elements since the Hamiltonian is a linear combination of 3 Pauli terms, so the measurement of each vector element requires 3 different quantum circuits. For example, the quantum circuits for measuring 
$\bra{\frac{\partial \psi}{\partial \theta_1}} H \ket{\psi} = \frac{\Delta E}{2} \bra{\frac{\partial \psi}{\partial \theta_1}} Z_1 \ket{\psi}  + V \bra{ \frac{\partial \psi }{\partial \theta_1}} X_1X_2 \ket{\psi} + V \bra{\frac{\partial \psi}{\partial \theta_1}} X_2 \ket{\psi}$ 
are displayed in Fig.~\ref{fig:V_1_Z1}, \ref{fig:V_1_X1X2} and \ref{fig:V_1_X2}. 

\begin{figure}[H]
    \centering
    \begin{tikzcd}
    \lstick{$\ket{0}$}&\gate{H}&\gate{R_z(\pi)}&\gate{X}&\qw&\ctrl{1}&\qw&\ctrl{1}&\gate{X}&\gate{H}&\meter{}\\
    \lstick{$\ket{0}$}&\qw&\qw&\gate{R_z(\theta_1)}&\gate[2]{R_{xx}(\theta_3)}&\gate{Z}&\gate[2]{R_{xx}(-\theta_3)}&\gate{Z}&\qw&\qw&\qw\\
    \lstick{$\ket{0}$}&\qw&\qw&\gate{R_x(\theta_2)}& &\qw& &\qw&\qw&\qw&\qw
    \end{tikzcd}
    \caption{Quantum circuit for computing $\mbox{Im} (\bra{\frac{\partial \psi}{\partial \theta_1}} Z_1 \ket{\psi})$.}
    \label{fig:V_1_Z1}
\end{figure}

\begin{figure}[H]
    \centering
    \begin{tikzcd}
    \lstick{$\ket{0}$}&\gate{H}&\gate{R_z(\pi)}&\gate{X}&\ctrl{1}&\ctrl{2}&\ctrl{1}&\gate{X}&\gate{H}&\meter{}\\
    \lstick{$\ket{0}$}&\qw&\qw&\gate{R_z(\theta_1)}&\gate{X}&\qw&\gate{Z}&\qw&\qw&\qw\\
    \lstick{$\ket{0}$}&\qw&\qw&\gate{R_x(\theta_2)} &\qw&\gate{X} &\qw&\qw&\qw&\qw
    \end{tikzcd}
    \caption{Quantum circuit for computing $\mbox{Im} (\bra{ \frac{\partial \psi }{\partial \theta_1}} X_1X_2 \ket{\psi})$.}
    \label{fig:V_1_X1X2}
\end{figure}

\begin{figure}[H]
    \centering
    \begin{tikzcd}
    \lstick{$\ket{0}$}&\gate{H}&\gate{R_z(\pi)}&\gate{X}&\ctrl{2}&\ctrl{1}&\gate{X}&\gate{H}&\meter{}\\
    \lstick{$\ket{0}$}&\qw&\qw&\gate{R_z(\theta_1)}&\qw&\gate{Z}&\qw&\qw&\qw\\
    \lstick{$\ket{0}$}&\qw&\qw&\gate{R_x(\theta_2)} &\gate{X} &\qw&\qw&\qw&\qw
    \end{tikzcd}
    \caption{Quantum circuit for computing $\mbox{Im} (\bra{\frac{\partial \psi}{\partial \theta_1}} X_2 \ket{\psi})$.}
    \label{fig:V_1_X2}
\end{figure}

\section{Mitigation of measurement errors}
Before we performed any quantum simulations with the IBM quantum computers, we calibrated the measurement noise with a calibration matrix $W$ on a daily basis. We first express the probabilities of obtaining the basis states $\ket{00}$, $\ket{01}$, $\ket{10}$ and $\ket{11}$ from a quantum circuit in a vector $C_{noisy}$.
Next we find the matrix $W$ that relates the probability vector $C_{noisy}$ with the ideal measurement outcome without errors $C_{ideal}$, i.e.  $C_{noisy} = W C_{ideal}$. 
To obtain the matrix $W$ for the 2-qubit case, we independently prepare each of the input states $\ket{00}$, $\ket{01}$, $\ket{10}$ and $\ket{11}$, and measure that probability for each outcome.

\section{Simulation Results with Noiseless Simulator}
\begin{figure}[H]
  \includegraphics[width=\linewidth]{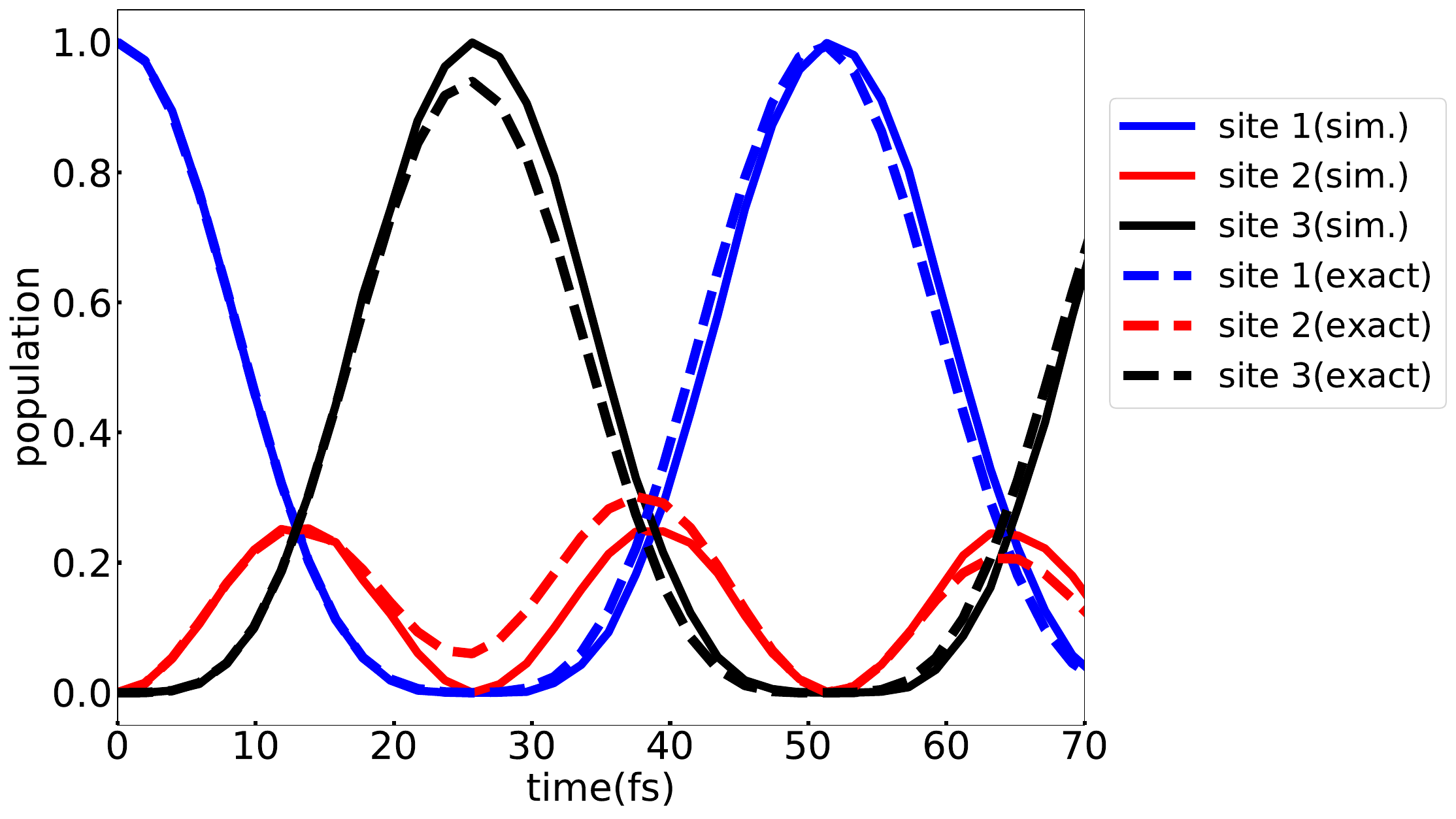}
    \caption{Population dynamics of the 4-site model obtained from VQA using the IBM Qiskit noiseless simulator (dashed lines). The solid lines are results from numerically exact calculations.}
    \label{fig:simulator}
\end{figure}

We repeat the VQA simulation of the 4-site system using the noiseless simulator in IBM Qiskit quantum simulator, and the results are shown in Fig.~\ref{fig:simulator}. The number of shots used is 8192. It can be seen that the noiseless simulator results are in good agreement with the exact calculations, indicating the error observed in Fig. 4a in the main text does not come from the choice of wavefunction ansatz or the finite number of shots. 

\section{Noisy Simulation of Full Hilbert Space Exciton Hamiltonian}
\begin{figure}[H]
  \includegraphics[width=\linewidth]{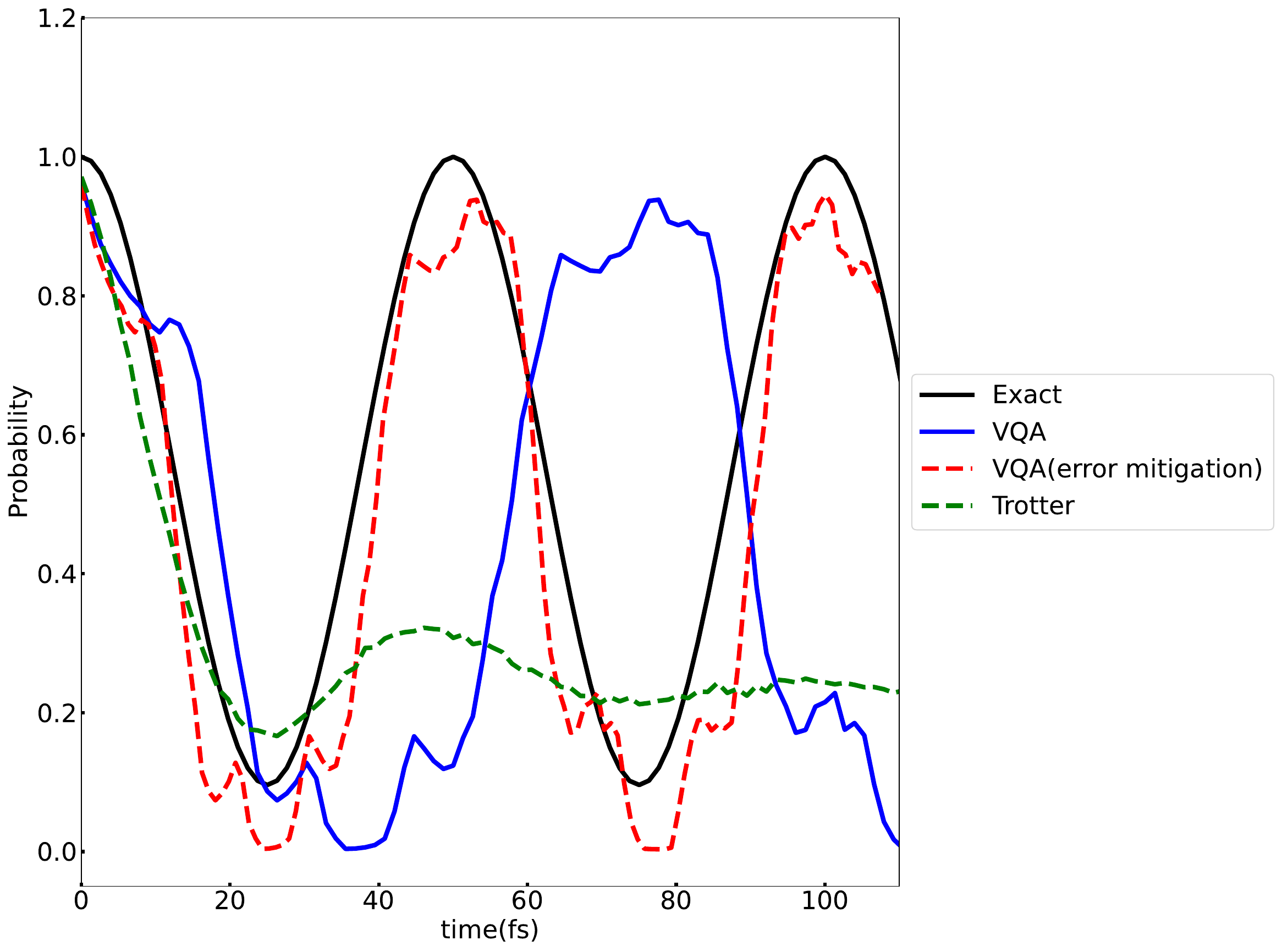}
    \caption{Dynamics of a pair of dimers with the full Hilbert space Hamiltonian. The results are obtained from exact diagonalization (black), VQA (blue), VQA with error mitigation (red dashed), and Trotter algorithm (green dashed).}
    \label{fig:fci_noisy}
\end{figure}

Here we present the exciton dynamics simulation results of a pair of dimers with the full Hilbert space Hamiltonian (see Eq.~(11) in the main text) using noisy quantum simulator. The noise parameters are extracted from the IBM’s \textit{ibmq$\_$bogota} quantum computer. 
The dimer is first initialized in state $\ket{01}$, i.e. the first molecule is initially excited whereas the second molecule is initially in the ground state. In Fig.~\ref{fig:fci_noisy} we investigate the excitation probability of the second molecule as a function of time, i.e. $|\langle 10 \ket{\psi}|^2$, since we are interested to study the transport of excitation from the first molecule to the second molecule. 
In our simulation, we use a Hamiltonian ansatz of single layer and time step of $1.3$fs.
From Fig.~\ref{fig:fci_noisy} we can see that the quantum dynamics obtained from the VQA with error mitigation (red) is more accurate than the results from the VQA without error mitigation (blue) and the Trotter algorithm (green), demonstrating that the proposed error mitigation method can also improve the dynamical simulation of systems with full Hilbert space Hamiltonians.

\section{Noisy Simulation of Transverse Field Ising Model}
\begin{figure}[H]
  \includegraphics[width=\linewidth]{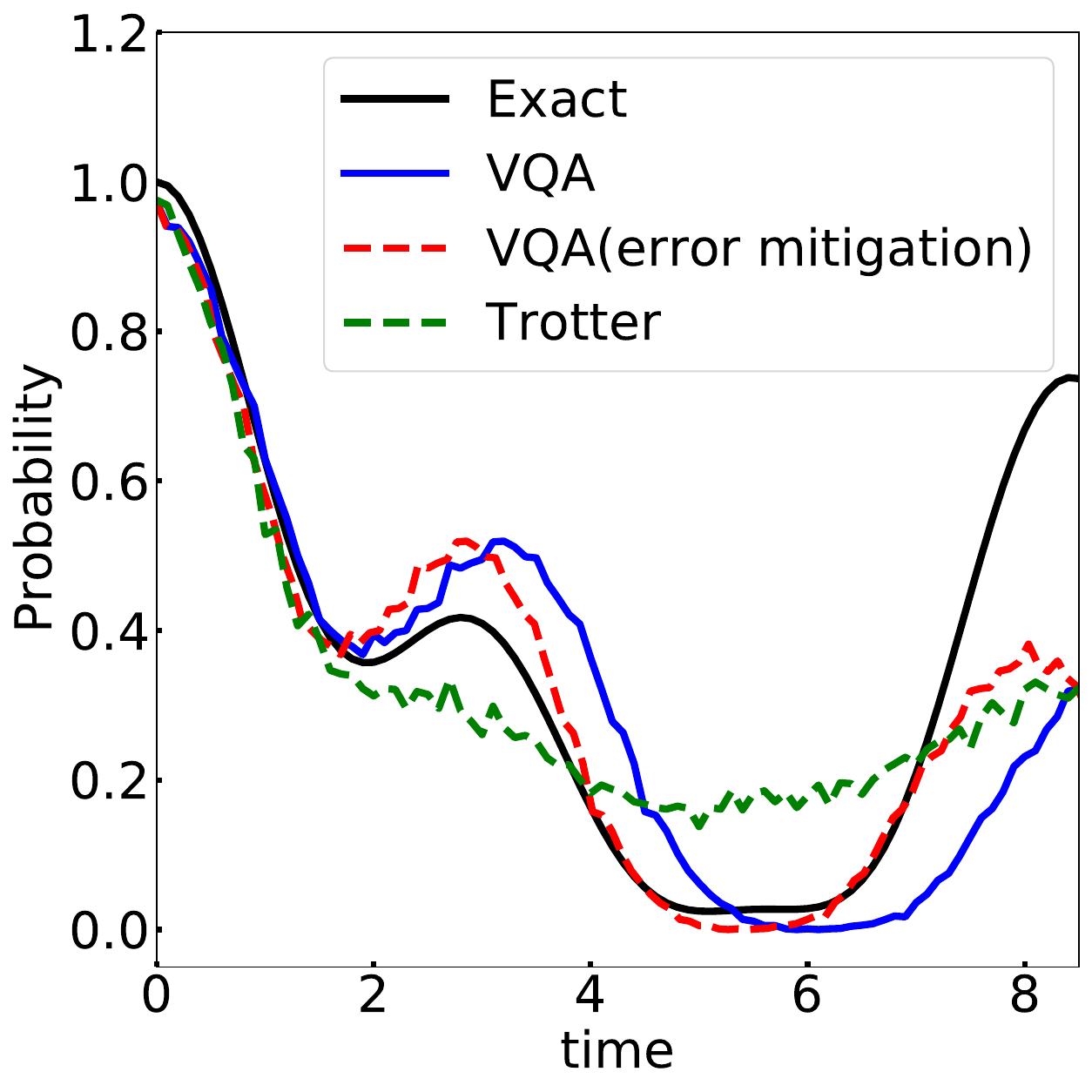}
    \caption{Dynamics of a TFI model obtained from exact diagonalization (black), VQA (blue), VQA with error mitigation (red dashed), and Trotter algorithm (green dashed).}
    \label{fig:TFI_noisy}
\end{figure}

In this section we simulate the dynamics of a 2-spin transverse field Ising (TFI) model with noisy simulator described in Sec.~\ref{sec:noise_model} to demonstrate  the generality of the proposed error mitigation method. 
The Hamiltonian of the TFI model is given by $H = h (\sigma_x^1 +  \sigma_x^2) + J \sigma_z^1  \sigma_z^2$ where $h$ is the magnitude of the transverse magnetic field and $J$ is the interaction strength. 
Here we use $h=0.5$, $J=0.5$, a noise level of $\lambda=0.01$, a Hamiltonian ansatz with depth 2 and time-step of $0.1$.
The system is initialized in state $\ket{00}$ and we investigate the probability of obtaining state $\ket{00}$ at time $t$, the results are shown in Fig.~\ref{fig:TFI_noisy}.
It can be seen that quantum dynamics obtained from the VQA with error mitigation (red) is more accurate than the results from the VQA without error mitigation (blue) and the Trotter algorithm (green). This suggests our error mitigation approach can be used for general quantum dynamical simulations. 

\section{Computational Details of the Full Hilbert Space Hamiltonian}

Following the work by Parrish and co-workers,\cite{Parrish2019b} the exciton Hamiltonian in the full Hilbert space of $N$ two-level chromophores can be expressed in terms of Pauli matrices, 
\begin{eqnarray} 
\hat{H}  = \mathcal{E} \hat I + \sum_{m=1}^N ( \mathcal{Z}_m \hat Z_m + \mathcal{X}_m \hat X_m ) + \sum_{m=1}^N \sum_{n<m} (\mathcal{XX}_{mn} \hat X_m \otimes \hat X_n + \label{eq:H_martinez} \\
\mathcal{XZ}_{mn} \hat X_m \otimes \hat Z_n + \mathcal{ZX}_{mn} \hat Z_m \otimes \hat X_n + \mathcal{ZZ}_{mn} \hat Z_m \otimes \hat Z_n ), \nonumber
\end{eqnarray}
where $\hat{I}$ is the identity matrix, and $\hat{X}_m$, $\hat{Y}_m$, and $\hat{Z}_m$ are the Pauli matrices associated with the $m$th chromophore. The coefficients in Eq. (\ref{eq:H_martinez}) are given by
\begin{eqnarray} 
\mathcal{E} = \sum_m S_m + \sum_{n<m} (S_m|S_n)\\
\mathcal{Z}_m = D_m + \sum_n (D_m|S_n) \\
\mathcal{X}_m = X_m + \sum_n (T_m|S_n) \\
\mathcal{XX}_{mn} = (T_m|T_n) \\
\mathcal{XZ}_{mn} = (T_m|D_n)\\
\mathcal{ZX}_{mn} = (D_m|T_n)\\ 
\mathcal{ZZ}_{mn} = (D_m|D_n).
\end{eqnarray}

In these expressions, $S_m$, $D_m$, and $X_m$ are determined by the matrix elements of one-body term, $\hat h$, associated with the $m$th chromophore,
\begin{eqnarray} 
S_m = [ ( 0_m | \hat h | 0_m ) + ( 1_m |  \hat h | 1_m ) ] /2 \\ 
D_m = [ ( 0_m | \hat h | 0_m ) - ( 1_m |  \hat h | 1_m ) ] /2\\
X_m = ( 0_m | \hat h | 1_m ),
\end{eqnarray}
where $| 0_m )$ and $| 1_m )$ are the ground and excited states of the $m$th chromophore. Note that following Ref. \citenum{Parrish2019b}, we use the chemists' notations for matrix elements. The terms of the form $(\cdots|\cdots)$ are associated with the two-body interaction term, $\hat v$, and symbols $|S_m)$, $|D_m)$, and $|T_m)$ denote
\begin{eqnarray} 
|S_m) = |0_m 0_m + 1_m 1_m )/2 \\ 
|D_m) = |0_m 0_m - 1_m 1_m )/2\\
|T_m) = |0_m 1_m).
\end{eqnarray}
For example, $(T_m | S_n) = (0_m 1_m | \hat v | 0_n 0_n + 1_n 1_n)/2$. Within the Born-Oppenheimer approximations, these coefficients are time-dependent, and their evolutions are governed by the classical MD simulation. In implementing this Hamiltonian for T2 molecules, we approximately set $( 0_m | \hat h | 0_m )=( 0_m | \hat h | 1_m )=0$, and $( 1_m | \hat h | 1_m )$ to be the lowest lying excited state energy of the $m$th T2 molecule from TDDFT calculations within the Tamm-Dancoff approximation (TDA). 

The coefficients for the two-body terms were computed using the approximation of dipole-dipole interaction, and the dipole position is taken to be the centers of mass of the T2 molecules. Ground-state, excited-state and transition dipoles were used for the terms involving $|0_m 0_m)$, $|1_m 1_m)$, and $|0_m 1_m)$, respectively. For example, 
\be
(0_m 0_m| \hat v | 0_n 1_n) \approx \frac{\vec{\mu}_{00,m} \cdot \vec{\mu}_{01,n} - 3(\vec{\mu}_{00,m} \cdot \hat{r}_{mn})(\vec{\mu}_{01,n} \cdot \hat{r}_{mn}) }{r_{mn}^3}, 
\ee
where $\vec{\mu}_{00,m}$ is the ground-state dipole of the $m$th T2, $\vec{\mu}_{01,n}$ is the transition dipole associated with the excitation on the $n$th T2, $\vec{r}_{mn}$ is the vector connecting the centers of mass of the two T2 molecules, $r_{mn} = |\vec{r}_{mn}|$, and $\hat{r}_{mn} = \vec{r}_{mn} / |\vec{r}_{mn}|$. These dipoles were obtained from DFT and TDA-TDDFT calculations. All the quantum-chemical calculations were performed at the level of CAM-B3LYP/6-31+G(d) using PySCF. 

\
\bibliographystyle{naturemag_noURL}
\bibliography{MyCollection}